\documentclass[12pt]{article}

\usepackage{times}

\usepackage{amsmath, setspace}
\usepackage{amssymb, xspace}

\usepackage{graphicx}

\usepackage{color} 
\usepackage{verbatim}

\usepackage{hyperref}
\usepackage{times}
\usepackage[scaled]{helvet}
\usepackage[round]{natbib} 
\usepackage[table,svgnames]{xcolor}
\usepackage{overpic}
\usepackage{array}
\usepackage{caption}
\newenvironment{sciabstract}{%
\begin{quote} \bf}
{\end{quote}}

\date{}

\definecolor{citecolor}{rgb}{0,0,0.4} 
\definecolor{linkcolor}{rgb}{0,0,0.4} 

\hypersetup{
colorlinks=true, 
citecolor=citecolor,
linkcolor=linkcolor,
urlcolor=linkcolor
}
\graphicspath{ {Figs/}			   
}

\newcommand{\fg}[1]{\textcolor{linkcolor}{Fig.~\ref{#1}}} 
\newcommand{\figs}[1]{\textcolor{linkcolor}{Figure E{#1}}} 
\newcommand{\supp}{Extended View Appendix\xspace	}

\newcommand*{\hvfont}{\fontfamily{phv}\selectfont}
\newcommand{\ina}[2]{ \begin{overpic}[width = .42\textwidth]{#1} \put(0,55){\large \bf \hvfont  #2}\end{overpic} }

\newcommand{\inbL}[2]{ \begin{overpic}[width = .88\textwidth]{#1} \put(0,75){\large \bf \hvfont #2}\end{overpic} }

\newcommand{\inaL}[2]{ \begin{overpic}[width = .98\textwidth]{#1} \put(0,56){\large \bf \hvfont  #2}\end{overpic} }

\newcommand{\BoxPlots}[2]{ \begin{overpic}[width = .48\textwidth]{#1} \put(-5,94){\large \bf \hvfont  #2}\end{overpic} } 
\newcommand{\BoxPlotsS}[2]{ \begin{overpic}[width = .24\textwidth]{#1} \put(-5,94){\large \bf \hvfont  #2}\end{overpic} }

\newcommand{\inYeast}[2]{ \begin{overpic}[width = .99\textwidth]{#1} \put(-4,98){\large \bf \hvfont  #2}\end{overpic} }
\newcommand{\inaSm}[2]{ \begin{overpic}[width = .48\textwidth]{#1} \put(-1,57){\large \bf \hvfont #2}\end{overpic} }
\newcommand{\inbSm}[2]{ \begin{overpic}[width = .43\textwidth]{#1} \put(-2,76){\large \bf \hvfont #2}\end{overpic} }

\newcommand{\BoxPlotsM}[2]{ \begin{overpic}[width = .4\textwidth]{#1} \put(-5,94){\large \bf \hvfont  #2}\end{overpic} }

\newcommand{\inb}[2]{ \begin{overpic}[width = .46\textwidth]{#1} \put(0,73){\large \bf \hvfont #2}\end{overpic} }
\newcommand{\IIms}[2]{ \begin{overpic}[width = .27\textwidth]{#1} \put(-4,105){\Large \bf \hvfont  #2}\end{overpic} }
\newcommand{\IIwb}[2]{ \begin{overpic}[width = .6\textwidth]{#1} \put(0,70){\Large \bf \hvfont #2}\end{overpic} }

\title{\Huge Differential stoichiometry among core \\ ribosomal proteins}

\author{ 
Nikolai Slavov$^{a,\ddagger}$,  
Stefan Semrau$^{b}$,
Edoardo Airoldi$^{a}$,
Bogdan A. Budnik$^{a}$,  \\
Alexander van Oudenaarden$^{b}$ 
\\ \\
\normalsize{$^{a}$Department of Statistics and FAS Center for Systems Biology,} \\
\normalsize{Harvard University, Cambridge, MA 02138, USA}\\
\normalsize{$^{b}$Hubrecht Institute, Royal Netherlands Academy of Arts and Sciences and } \\ \normalsize{University Medical Center Utrecht, Uppsalalaan 8, 3584 CT, Utrecht, The Netherlands}\\
\\ 
{$^\ddagger$To whom correspondence should be addressed: \href{mailto:nslavov@alum.mit.edu}{nslavov@alum.mit.edu} }
}

\begin{document}
\maketitle

\vspace{2cm}

\begin{spacing}{1.45}

\newpage
\begin{sciabstract}
Understanding the regulation and structure of ribosomes is essential to understanding protein synthesis and its deregulation in disease.  While ribosomes are believed to have a fixed stoichiometry among their core ribosomal proteins (RPs), some experiments suggest a more variable composition. Testing such variability requires direct and precise quantification of RPs. We used mass-spectrometry  to  directly quantify RPs across monosomes and polysomes of mouse embryonic stem cells (ESC) and budding yeast. Our data show that the stoichiometry among core RPs in wild-type yeast cells and ESC depends both on the growth conditions and on the number of ribosomes bound per mRNA. Furthermore, we find that the fitness of cells with a deleted RP-gene is inversely proportional to the enrichment of the corresponding RP in polysomes. Together, our findings support the existence of ribosomes with distinct protein composition and physiological function.   
\end{sciabstract}

%
\section*{Introduction} 
Ribosomes catalyze protein translation but have only a few characterized roles in regulating translation \citep{mauro2002ribosome_filter, xue2012specialized}.  Rather, the most--studied molecular regulatory mechanisms of translation are mediated by eukaryotic initiation factors, RNA binding proteins, and microRNAs \citep{Sonenberg_1999_eif4,  microRNAs_Ferrell_Brown_2009, sonenberg2009regulation, sonenberg2012_microRNAs}. 
The characterized catalytic role of the ribosomes corresponds well to the model of the ribosome as a single complex with a fixed stoichiometry: 4 ribosomal RNAs and 80 core RPs \citep{warner1999economics, Xribo2010crystal, Xribo2011crystal_atomic}, some of which are represented by several paralogous RPs. Despite the longstanding interest in ribosome structure and function, the exact stoichiometry and possible heterogeneity of the ribosomes have been challenging to measure directly \citep{weber1972stoichiometric, westermann1976stoichiometry, hardy1975stoichiometry}. Such measurements are enabled by modern quantitative mass spectrometry (MS); reviewed by \citet{Aebersold2012review}. Indeed, MS has transformed our understanding of protein complexes, such as proteasomes \citep{Huang_2007_proteasome} and nuclear pore complexes \citep{npc_2013_ori_beck, npc_2013_beck}, by demonstrating variability among their protein  subunits.

Studies of eukaryotic ribosomes \citep{RPL13_Paul_Fox_2003, galkin2007roles, komili2007functional, RPL38_kondrashov2011ribosome, Sonenberg2011Preview, horos2012ribosomal, lee2013ribosome, tiruneh2013global} have demonstrated that ($i$) genetic perturbations to the core RPs specifically affect the translation of some mRNAs and not others and ($ii$) mRNAs coding for core RPs are transcribed, spliced, and translated differentially across physiological conditions \citep{ramagopal1981regulation, ramagopal1990induction, brauer_2008, parenteau2011introns, Slavov_2009, Slavov_eth_grr, rpl22_repressing2013, Slavov_exp, jovanovic2015dynamic}.       
These results suggest the hypothesis \citep{mauro2002ribosome_filter, gilbert2011functional, xue2012specialized} that, depending on the tissue type and the physiological conditions, cells can alter the stoichiometry among the core RPs comprising the ribosomes and thus in turn alter the translational efficiency of distinct mRNAs. Alternatively, differential RP-expression can reflect extra ribosomal functions of the RPs \citep{RPL13_Paul_Fox_2003, wool1996extraribosomal, warner2009common}. Furthermore, polysomes (multiple ribosomes per mRNA) from different cell--lines have similar core RP stoichiometries \citep{riboproteome_2013_Pandolfi}. Thus, the variable RP stoichiometry in the ribosomes of wild-type cells that is suggested by the ribosome specialization hypothesis  remains unproven.

\section*{Results}
\subsection*{Differential stoichiometry among core RPs in mouse ESCs}
To measure whether the stoichiometry among RPs can vary, we quantified RP levels in monosomes and polysomes from exponentially growing mouse embryonic stem cells (ESC). The ribosomes were isolated by velocity--sedimentation in sucrose--gradients (\fg{mouse}A), and the proteins from individual fractions were digested to peptides, labeled with tandem mass tags (TMT), and quantified on Orbitrap Elite based on the MS2 intensities of the TMT reporter ions; see \supp.
 The monosomal sample was quantified in two replicas (1a and 1b), and the results indicate very high reproducibility ($\rho =  0.92$; \fg{mouse}B). To control for protease and peptide biases, the proteins from each analyzed sucrose fraction were digested either by trypsin (T) or by lys--C (L) and peptides from each digestion quantified independently. 
Because of the different specificity of trypsin and lys-C, most RP peptides (1058) were identified and quantified only in the trypsin or only in the lys--C digestion, while only 269 peptides were identified and quantified in both digestions. Thus, peptide-specific biases (such as co-isolation interference) cannot be shared between the two digestions.     

The measured levels of a unique peptide (a peptide present in a single RP) reflect the levels of the corresponding RP, post--translational modifications (PTMs) of the peptide (if any), and measurement error. We quantify on average ten distinct RP--peptides per RP (\figs{1}A), and the levels of these peptides allow both the estimation of the RP levels and the consistency of these estimates. To depict both the estimates and their consistency, we display the full distributions of relative levels of all peptides unique to an RP as boxplots in \fg{mouse}C, D. 
The RP levels across the sucrose gradient (estimated as the median of the levels of unique peptides) indicate that some RPs are enriched in monosomes (\fg{mouse}C), while other RPs are enriched in polysomes (\fg{mouse}D). 
This enrichment is substantially higher than the measurement noise, consistent across replicas and across distinct peptides, and highly statistically significant at false discovery rate (FDR) $< 10^{-6}$.  
The relative levels of all RPs with quantified unique peptides are displayed in \fg{mouse2} to illustrate the global pattern of RP levels across monosomes and polysomes. This pattern shows more RPs whose variability is consistent across replicas and enzymatic digestions. In contrast, the levels of RPs buried in the core of the ribosomes remain constant, with estimates fluctuating within the tight bounds of the measurement noise, \fg{mouse2}. This fixed stoichiometry among RPs constituting the ribosomal core suggests that even ribosomes lacking some surface RPs likely have the same core structure; see \supp.


In principle, if only a few peptides are quantified per RP, the measured peptide variability might reflect reciprocal variability in corresponding PTM isoforms (if any) across the sucrose gradients, e.g., the unmodified isoform is enriched in monosomes and a phosphorylated isoform is enriched in polysomes. Such variable PTM isoforms (if any) are very interesting but cannot explain the data for an RP quantified by dozens of peptides spanning the protein length and indicating highly--consistent fold--changes across the sucrose gradient; see \fg{mouse}, \figs{1}, and \supp.  
  We further tested the RP variability with an independent method, Western blots, and in another strain of mouse ESC. Consistent with the MS data in \fg{mouse2}, the Western Blots data (\figs{3}) indicate that Rps29 and Rps14 are enriched in polysomes, Rpl11 is enriched in monosomes, and Rpl32 does not change beyond the measurement noise.

\subsection*{Differential stoichiometry among core RPs in yeast}  							
Having found differential stoichiometry among mouse RPs, we sought to further explore ($i$) whether such ribosome heterogeneity is conserved to budding yeast and ($ii$) whether the RP stoichiometry can change with growth conditions and metabolic state.   
To this end, we employed sucrose gradients to separate the ribosomes from yeast cells grown in minimal media with either glucose or ethanol as the sole source of carbon and energy \citep{Slavov_exp};  see \supp. Consistent with previous observations that the type and the concentration of the carbon source influence the ratio of monosomes to polysomes \citep{ashe2000glucose, castelli2011glucose, pavan2014wendy}, the ratio of monosomes to polysomes in our yeast cells grown in $0.4 \: \%$ ethanol (\fg{yeast}A) or in $0.2 \: \%$ glucose (\fg{yeast}B) is higher than is typically observed for yeast grown in rich media containing $2 \: \%$ glucose.  As in mouse, some RPs are enriched in monosomes (\fg{yeast}C) and others in polysomes (\fg{yeast}D, E). This enrichment is reproducible  (correlation between replicas $\rho = 97$; \fg{yeast}F) and consistent across independent unique peptides whose levels are shown as boxplot distribution in  \fg{yeast}C-D.


The pattern of relative RP levels shown in \fg{yeast}C-E indicates that RP stoichiometry depends on two factors: on the number of ribosomes per mRNA (as in mouse) and on the carbon source in the growth media; the RP levels that are higher in glucose compared to ethanol also tend to increase with the number of ribosomes per mRNA (\fg{yeast}C-E). Furthermore, the ratios between the polysomal and monosomal levels of yeast RPs correlate to the corresponding ratios for their mouse orthologs  (\fg{yeast}G; p-value $<0.03$), suggesting that the RP-stoichiometry differences between monosomes and polysomes are conserved across yeast and mouse.

Many yeast RPs are represented by two highly-homologous paralogs, and we explored whether the exchange among paralogs (one paralog substituting for the other) can account for the measured differential stoichiometry in \fg{yeast}E. 
The levels of paralogs localized on the surface of the ribosome, such as Rpl17a and Rpl17b, are positively correlated and thus inconsistent with paralog--exchange across the analyzed ribosomes (\fg{yeast}E).
In contrast, RPs embedded deep in the core of the ribosomes either remain constant (the estimated fluctuations of their levels are within errorbars) or their paralogs exchange (e.g., the levels of Rpl37a and Rpl37b are anticorrelated; see \fg{yeast}E), indicating that each ribosome has a copy of Rpl37. In general, the RPs whose levels differ the most
among the different fractions are located on the surface of the yeast ribosomes, as can be seen from their 3D color–coded rendition in the Supporting PDB files and Movie 1.

\subsection*{RP enrichment in polysomes correlates to fitness}
Next, we tested the variability among RPs and its phenotypic consequences by independent fitness measurements.  
Our observation that the RP stoichiometry depends on the number of ribosomes bound per mRNA parallels measurements of higher translational activity of polysomes compared to monosomes \citep{warner1963multiple, goodman1963mechanism}; some studies have even reported that the translational activity per ribosome increases with the number of ribosomes bound per mRNA \citep{noll1963ribosomal, wettstein1963ribosomal}, but this finding has not been widely reproduced. 
 We therefore hypothesized that genetic deletions of RPs enriched in the more active ribosomes -- as compared to RPs enriched in less active ribosomes -- may result in a larger decrease of the translation rate and thus lower fitness.  To test this hypothesis, we computed the correlation (\fg{fitness}A) between the fitness of yeast strains with single RP--gene deletions \citep{qian2012genomic}  and the corresponding relative RP levels measured in the tetra-ribosomal fraction (4 ribosomes per  mRNA).
Consistent with our hypothesis, the fitness of strains lacking RP--genes is inversely proportional to the relative levels of the corresponding RPs in the tetra-ribosomes (\fg{fitness}A). Extending this correlation analysis to the RP--levels in all sucrose fractions shown in \fg{yeast}E results in a correlation pattern (\fg{fitness}B) that further supports our hypothesis by showing the opposite dependence for fractions with fewer ribosomes per mRNA:  the fitness of strains lacking RP--genes is proportional to the relative levels of the corresponding RPs in fractions with fewer ribosomes per mRNA (\fg{fitness}B). This correlation pattern holds both for ethanol and for glucose carbon sources. 
 To mitigate possible artifacts in the fitness data due to potential chromosome duplications in the deletion strains, we computed the correlations between the RP--levels and the fitness of the corresponding RP--deletion strains only for RPs without paralogs (thus unlikely to be affected by chromosome duplication) and found much higher magnitudes of the correlations (\fg{fitness}A, B). This result suggests that the differential RP stoichiometry is not limited to paralogous RPs substituting for each other. 

To further explore the functional significance of the differential RP stoichiometry, we examined whether polysome--enriched RPs are preferentially induced at higher growth--rates. We previously found that the degree of growth-rate-dependent transcriptional induction varies significantly across RPs \citep{brauer_2008, Slavov_eth_grr, Slavov_emc}. We quantified the growth-rate responses of RPs by regressing their mRNA levels on growth-rates and computing growth rate-slopes. 
The magnitudes of RP growth--rate slopes range from positive (mRNA levels increase with increasing growth rate) to negative (mRNA levels decrease with increasing growth rate), see \figs{4}.   
Analogously to our fitness analysis (\fg{fitness}A), we correlated the growth-rate slopes to the relative RP levels from \fg{yeast}E. Consistent with our hypothesis, the correlation pattern (\fg{fitness}C) indicates that the higher the growth--rate slope of a RP, the higher its enrichment in sucrose-fractions corresponding to increasing numbers of ribosomes per mRNA.

We extended our fitness analysis from yeast to mouse using the published depletion data from CRISPR knockouts in human ESC  \citep{shalem2014genome}; see \supp. We used  BLAST to identify the closest mouse orthologs of each human RP with depletion data (\fg{fitness}D), and correlated the fitness of human ESC lacking the human RP orthologs to the RP levels across sucrose fractions that we measured, \fg{mouse2}. 
The correlation pattern (\fg{fitness}E) is  similar to the one in yeast (\fg{fitness}A-C)  and highly significant, FDR $< 0.1 \% $. This pattern indicates that the fitness of ESC lacking RP-genes is directly proportional to the relative RP levels in monosomes and inversely proportional to the relative RP levels in polysomes.  The  magnitude of this inverse proportionality increases with the number of ribosomes per mRNA  (\fg{fitness}E), consistent with our hypothesis. The fact that the fitness of human ESC lacking RPs correlates significantly to the levels of the corresponding mouse orthologous RPs suggests that the differential RP stoichiometry and its biological functions are likely conserved across mouse and human. The magnitude of this correlation increases when the correlation is computed based only on the orthologs whose sequences are over $80\%$ identical between mouse and human (\fg{fitness}E), providing further evidence for the conserved fitness consequences of the altered RP stoichiometry.

\section*{Discussion}
For decades, the ribosome has been considered the preeminent example of a large RNA--protein complex with a fixed stoichiometry among the constituent core RPs \citep{warner1999economics, Xribo2010crystal, Xribo2011crystal_atomic}. However, the direct and precise measurements of RP--levels required to support this view have been very challenging. Prior to our work, the most direct and precise quantification of RP stoichiometry that we know of is based on measuring the radioactivity from RPs labeled with $^{14}C$ or $^{3}H$ and separated on   2D--gels.
Some of these studies \citep{weber1972stoichiometric, westermann1976stoichiometry} achieved  very high precision (standard error $< 10\;\%$) and reported over 2--fold deviation from $1:1$ stoichiometry for multiple RPs. Other studies of prokaryotic ribosomes  \citep{hardy1975stoichiometry} achieved lower precision, and the deviation from $1:1$ stoichiometry was within the experimental error of the measurements.  The results reported in ref.~\citep{weber1972stoichiometric, westermann1976stoichiometry, hardy1975stoichiometry} are all consistent with our findings, albeit our measurements are limited to eukaryotic ribosomes. This prior work and our measurements reflect population--averages across a heterogeneous pool of ribosomes and thus likely underestimate the magnitude of the variability among RP stoichiometries.

A simple mechanism that may account for our observations is that the rates of translation initiation and elongation depend on the RP composition. Ribosomes whose RP composition corresponds to higher ratios between the initiation and the elongation rates are likely to be found in fractions with multiple ribosomes per mRNA. Conversely, ribosomes whose RP composition corresponds to lower ratios between the initiation and the elongation rates are likely to be found in fractions with fewer ribosomes per mRNA. Indeed, increased growth--rate on glucose carbon-source that we find associated with altered RP stoichiometry has been previously reported to be associated with faster elongation rates \citep{bonven1979peptide, young1976polypeptide}.

Our data show that the stoichiometry among RPs varies across ribosomes from different sucrose fractions. However, velocity sedimentation in sucrose gradients is unlikely to perfectly separate ribosomes based on their RP composition. For example, short mRNAs and the ribosomes translating them can be found only in the fractions containing few ribosomes per mRNA regardless of the efficiency of translation and the RP-composition of the ribosomes \citep{arava2003genome}. Similarly, even the most highly translated mRNA that is likely to be translated by polysome-type ribosomes will go through a stage when only a single ribosome is loaded and thus will be found in the monosomal fraction. Other factors may also contribute to the mixing of different ribosomes in each sucrose fraction, including variation in the mRNA length, any degree of ribosome run-off, and mRNA shearing during sample handling, if any. None of these factors, however, is likely to artifactually contribute to the differential RP stoichiometry that we observe. Rather, the presence of ribosomes with different RP compositions in the same sucrose fraction would average out and decrease the differences, resulting in underestimation of the RP variability.   

The conserved difference between monosomal and polysomal ribosomes (\fg{yeast}G) raises the question about the activity of monosomes, especially given the lower estimates for their transnational activity \citep{warner1963multiple, wettstein1963ribosomal}. The RP levels in \fg{yeast}E indicate that the RP composition of trisomes in ethanol is more similar to the composition of monosomes than to tetrasomes. This observation shows that monosomes may have similar RP composition to polysomes, suggesting that the RP composition of monosomes is not necessarily indicative of a non--functional state. 

The correlations between RP--composition and fitness can be explained by the expectation that the higher the translational activity of a ribosome, the higher the fitness cost of its perturbation in rapidly growing stem cells. The key factor required for this expectation is the variable ribosomal composition that we measured. The variable RP stoichiometry in the absence of external perturbations suggests that cells use variable RP composition of their ribosomes as a regulatory mechanism of protein translation. One such example might be the preferential transcriptional induction of polysome-enriched RPs at higher growth rates (\fg{fitness}C).

Variable mammalian RPs, such as Rps4x, Rps14, Rps20, Rpl5, Rpl10, and Rpl27, directly bind mRNAs \citep{castello2012insights, kwon2013rna}, and this binding might mediate translational regulation as previously suggested \citep{mauro2002ribosome_filter, RPL13_Paul_Fox_2003, xue2015rna}. Furthermore,  deletions or overexpressions of many of the variable RPs (\fg{mouse}B) have well characterized phenotypes both in development and in cancer. For example, the knockdown or haploinsufficiency  of the polysomally enriched Rps19 (\fg{mouse}B) causes Diamond Blackfan anemia by selectively affecting the synthesis of some proteins but not of others \citep{horos2012ribosomal}.
 Interestingly, our data indicate that RPs that are frequently mutated in cancers, such as Rpl5 and Rpl10 \citep{de2013exome, lawrence2014discovery}, are enriched in the monosomes (\fg{mouse}A and \fg{mouse2}).  
Conversely, RPs whose (over)--expression promotes cancer, such as Rpl30, Rps20, and Rpl39 \citep{de2006medulloblastoma, dave2014targeting},  are enriched in the polysomes (\fg{mouse}B and \fg{mouse2}). One interpretation, among others, of these data is that loss of function of monosomally--enriched RPs or overexpression of polysomally--enriched RPs might promote protein synthesis and cancer cell growth.


\section*{Materials and Methods}
All yeast experiments used a prototrophic diploid strain (DBY12007) with a S288c background and wild type HAP1 alleles \citep{hickman2007heme}. We grew our cultures in a bioreactor (LAMBDA Laboratory Instruments) using minimal media with the composition of yeast nitrogen base (YNB) and supplemented with $2g/L$ D-glucose. Before inoculation, the reactor was filled with 2L of minimal media and warmed up to a working temperature of $30^oC$.

Mouse embryonic stem cells (E14 $10^{th}$ passage) were grown as adherent cultures in 10 cm plates with 10 ml  DMEM/F12 media supplemented with 10 \% knockout serum replacement, nonessential amino acids (NEAA supplement), 0.1 mM $\beta$--mercapto--ethanol, 1 \% penicillin and streptomycin, leukemia inhibitory factor (LIF; 1,000 U LIF/ml), and $2i$ (GSK3$\beta$ and Mek 1/2 inhibitors).

Both yeast and mouse embryonic stem cells were lysed by vortexing for 10 min with glass beads in cold PLB (20 mM HEPES--KOH at pH 7.4, 1 \% Triton X--100, 2 mM Magnesium Acetate, 100 mM Potassium Acetate, 0.1 mg/ml cycloheximide, and 3 mM DTT). The crude extracts obtained from this lysis procedure were clarified by centrifugation. The resulting supernatants were applied to linear 11 ml sucrose gradients ($10 \; \% - 50 \; \%$) and spun at 35,000 rpm in a Beckman SW41 rotor either for 3 hours (for yeast samples) or for 2.5 hours (for mouse samples). Twelve fractions from each sample were collected using a Gradient Station. More details are available in the \supp.


\section*{Data availability}
The raw and processed MS data, and 3D structures of the yeast ribosomes color--coded  according to the RP levels from Fig.~3E  are enclosed with this submission and can also be found at:\\
\href{http://alum.mit.edu/www/nslavov/}{http://alum.mit.edu/www/nslavov/}




\subsection*{Acknowledgments}
We thank J.~Cate and N.~Lintner for helping us colorcode the variability of RPs on the 3D structure of the yeast ribosomes, P.~Vaidyanathan for help with the sucrose gradients, R.~Robertson for technical assistance, and M.~Jovanovic, S.~Kryazhimskiy, W.~Gilbert, P.~Vaidyanathan, Y.~Katz, G.~Frenkel, D.~Mooijman, J.~Alvarez, D.~Botstein, and A.~Murray for  discussions and constructive comments. This work was funded by a grant from the National Institutes of Health to A.v.O. (R01-GM068957) and Alfred P Sloan Research Fellowship to E.M.A.
%

\subsection*{Author Contributions}
N.S designed research and experimental procedures. N.S., S.S., and B.B.~performed experiments and collected data. N.S., A.v.O.,~and E.A.~analyzed the data. N.S., S.S., and A.v.O.~discussed the results and wrote the manuscript.

\subsection*{Conflict of Interest}
The authors declare no conflict of interest.

\clearpage

\section*{Figure Captions}

\begin{figure}[h!]
\caption{ {\bf The stoichiometry among core RPs in mouse ribosomes depends on the number of ribosomes per mRNA.} 
({\bf A}) Sucrose gradients allow separating ribosomes that are free or bound to a single mRNA (monosomes, depicted in black) from multiple ribosomes bound to a single mRNA (polysomes, depicted in blue).  The absorbance at 254 nm reflects RNA levels, mostly ribosomal RNA. The vertical dashed lines indicate the boundaries of the collected fractions. 
 ({\bf B}) Replica MS measurements of the monosomes (1a and 1b) indicate reproducible estimates  for RP enrichment in polysomes. 
({\bf C}) Some RPs are enriched in monosomes and others in polysomes ({\bf D}). The relative levels of each RP are quantified as the median levels of its unique peptides, and the probability that the RP levels do not change across the quantified fractions is computed from ANOVA (indicated at the top). 
The distributions of levels of all unique peptides from trypsin (left panels) and from lys-C (right panels) digestions are juxtaposed as boxplots to 
depict the consistency of the estimates across proteases, different peptides, and experiments.
 To correct for variation in the total amount of ribosomes present in each fraction, the mean of the intensities of all RP peptides was normalized to 1 for each fraction.  
On each box, the central line is the median, the edges of the box are the $25^{th}$ and $75^{th}$ percentiles, and the whiskers extend to the most extreme data points.       
}
\label{mouse}
\end{figure}


\begin{figure}[h!]
\caption{ {\bf Global pattern of differential stoichiometry among mouse RPs across sucrose gradients.}   
The relative levels of core RPs 
in monosomes and polysomes were quantified by MS and found to vary  depending on the number of ribosomes bound per mRNA. The measurement noise was estimated by $(i)$ replica quantification of the monosomal fraction (by using different tandem-mass-tags reporter ions, 126 or 131) and by $(ii)$ estimating RP levels separately using either trypsin (T) or lys-C (L)  digestion, as indicated at the bottom of each column. The $\log_2$ levels of each RP are shown relative to their mean.  See \supp and \figs{1} for more details and error estimates. 
}
\label{mouse2}
\end{figure}

\begin{figure}[h!]
\caption{ {\bf The stoichiometry among core RPs in yeast ribosomes depends both on the number of ribosomes per mRNA and on the physiological condition.}
Ribosomes from either ethanol ({\bf A}) or glucose  ({\bf B}) grown yeast were separated by velocity sedimentation in sucrose gradients. Depiction is as in \fg{mouse}A.   
 ({\bf C}) Rpl35b is enriched in monosomes ($p < 10^{-3}$) and in ethanol carbon source ($p < 10^{-3}$).  Depiction is as in \fg{mouse}. The p value at the top is computed from ANOVA and quantifies the probability of observing the variability of Rpl35b peptides by chance.   
 ({\bf D}) Rpl26a is enriched in polysomes ($p < 10^{-9}$) and in glucose carbon source ($p < 10^{-4}$).     
 ({\bf E}) Levels of core RPs in the sucrose fractions estimated from their unique peptides quantified by MS. The RP levels vary depending on the carbon source (glucose or ethanol) and on the number of ribosomes bound per mRNA, indicated at the top. Monosomes from ethanol grown yeast were quantified in two biological replicas (first two columns). The $\log_2$ levels of each RP are shown relative to their mean. See Supporting Movie 1 and PDB files for color-coded
depiction of these data on the 3D structure of the yeast ribosome.
 ({\bf F})  The RP fold--changes between the tetrasomes of yeast grown in glucose carbon source and the monosomes of yeast grown in ethanol carbon source are highly reproducible. The ethanol samples were collected and processed independently and compared to the glucose tetrasomes. 
  ({\bf G}) The $\log_2$ ratios between polysomal and monosomal levels of mouse RPs are plotted against the corresponding $\log_2$ ratios of their orthologous yeast RPs. The significant (p-value $<0.03$) positive correlation between these ratios suggests that the RP-stoichiometry differences are conserved across yeast and mouse. The plot includes all orthologous RP pairs with over $65\%$ sequence identity between yeast and mouse.           
}
\label{yeast}
\end{figure}

%

\begin{figure}[h!]
\caption{ {\bf The relative levels of RPs across monosomes and polysomes correlate significantly to the fitness of yeast and mammalian cells lacking the genes encoding these RPs.} 
({\bf A}) The fitness of RP--deleted yeast strains \citep{qian2012genomic} is inversely proportional (p--value $<4 \times 10^{-3}$) to the relative levels of the corresponding RPs in tetrasomes from yeast growing on ethanol carbon source. The RPs without paralogs are marked with red squares. 
 ({\bf B})  Extension of the analysis in panel (A) to all sucrose fractions:  correlations between the relative RP levels from \fg{yeast}E and the fitnesses of strains lacking the corresponding RP genes \citep{qian2012genomic}. The correlations are shown either for all quantified RPs or only for RPs without paralogs. 
 ({\bf C})  Correlations between the relative levels of the RPs from \fg{yeast}E and the their transcriptional growth--rate responses (slopes). The growth-rate slopes were previously computed by regressing ($R^2 > 0.87$) the levels of mRNAs in glucose-limited steady-state cultures of yeast against the growth-rates of the cultures \citep{brauer_2008, Slavov_eth_grr}. 
({\bf D}) Distribution of sequence identity between human RPs and their closest mouse orthologs; the sequences and annotations for RPs are from Swiss--Prot.     
({\bf E})  Extension of the analysis for yeast in panels (A-B) to mouse: correlations between the relative levels of mouse RPs from \fg{mouse2} and the fitness of human ESC lacking the corresponding  human ortholog \citep{shalem2014genome}. The correlations are shown either for all quantified RPs or only for RPs whose sequence identity between mouse and human exceeds $80 \; \%$. All error bars are standard deviations from bootstrapping.
}
\label{fitness}
\end{figure}
\clearpage

\setcounter{figure}{0}  \renewcommand{\thefigure}{ E\arabic{figure}}

\section*{Expanded View Figure Captions}

\begin{figure}[h!]
\caption{ {\bf Multiple unique peptides per RP provide consistent fold--change estimates for most RPs.} 
 ({\bf A}) Number of unique peptides quantified per mouse RP digested by trypsin.
 ({\bf B}) Distribution of coefficients of variation (CVs) of the measured fold--changes for mouse RPs digested by trypsin. 
 ({\bf C}) Number of unique peptides quantified per mouse RP digested by lys-C.
 ({\bf D}) Distribution of coefficients of variation (CVs) of the measured fold--changes for mouse RPs  digested by lys-C.  
 ({\bf E}) Number of unique peptides quantified per yeast RP digested by trypsin.
 ({\bf F}) Distribution of coefficients of variation (CVs) of the measured fold--changes for yeast RPs digested by trypsin.  
 The CVs for each RP quantify the consistency of fold-changes for that RP estimated from all quantified unique peptides whose amino acid sequences are found only in the RP and no other protein in the proteome. CVs are estimated as the standard deviation ($\sigma$) of the fold-changes of unique peptides mapping to the same RP over the corresponding mean ($\mu$) and are thus computed only for RPs that have multiple unique peptides. 
 ({\bf D}) Distribution of CVs of the measured fold-changes for mouse RPs. 
}
\label{artifacts}
\end{figure}
\vspace{1cm}

%

\begin{figure}[h!]
\caption{ {\bf RPs are over 100 fold more abundant that ribosome biogenesis proteins in the sucrose gradients. } 
Distributions of iBAQ scores for RPs and for ribosome biogenesis proteins.  The iBAQ score of a protein estimates its absolute level based on the number of unique peptides for that proteins and their corresponding integrated precursor areas.  The levels of ribosome biogenesis proteins likely reflect the levels of the 90S preribosomal particles in our sucrose gradients.  
}
\label{iBAQ}
\end{figure}
\vspace{1cm}

\begin{figure}[h!]
\caption{ {\bf Comparison of relative RP quantification by MS and Western Blots. } \\
 ({\bf A}) Polysomal enrichment of RPs quantified by MS.
 ({\bf B}) Polysomal enrichment of RPs quantified by Western blots. RPs were quantified by Western blots in monosomes and polysomes from high passage--number E14 mouse ESCs. Rpl32 was used as a loading control and the boxplots summarize data from 9 ratios for each quantified RP.
}
\label{WB}
\end{figure}
\vspace{1cm}

\begin{figure}[h!]
\caption{ {\bf Different RPs have different magnitudes of their growth-rate responses, i.e., transcriptional induction or repression with increasing growth rate.} \\
The heatmap displays transcript levels of RPs in yeast cultures growing at steady-state in glucose-limited minimal media at the growth-rates ($\mu$) indicated by the bars on the top. To emphasize the growth-rate trends, the mRNA levels of each RP are displayed on a $log_2$ scale relative to their mean across all six growth rates. The RPs are sorted by their growth-rate slopes to emphasize the variability of their slopes, from highly positive to negative. All data are from \citep{Slavov_eth_grr, brauer_2008}.  
}
\label{Grr_RPs}
\end{figure}
\vspace{1cm}

\clearpage

\section*{Figure 1}
\vspace{8mm}
\begin{figure}[h!]
\begin{minipage}{0.45\textwidth} 
 	\inaL{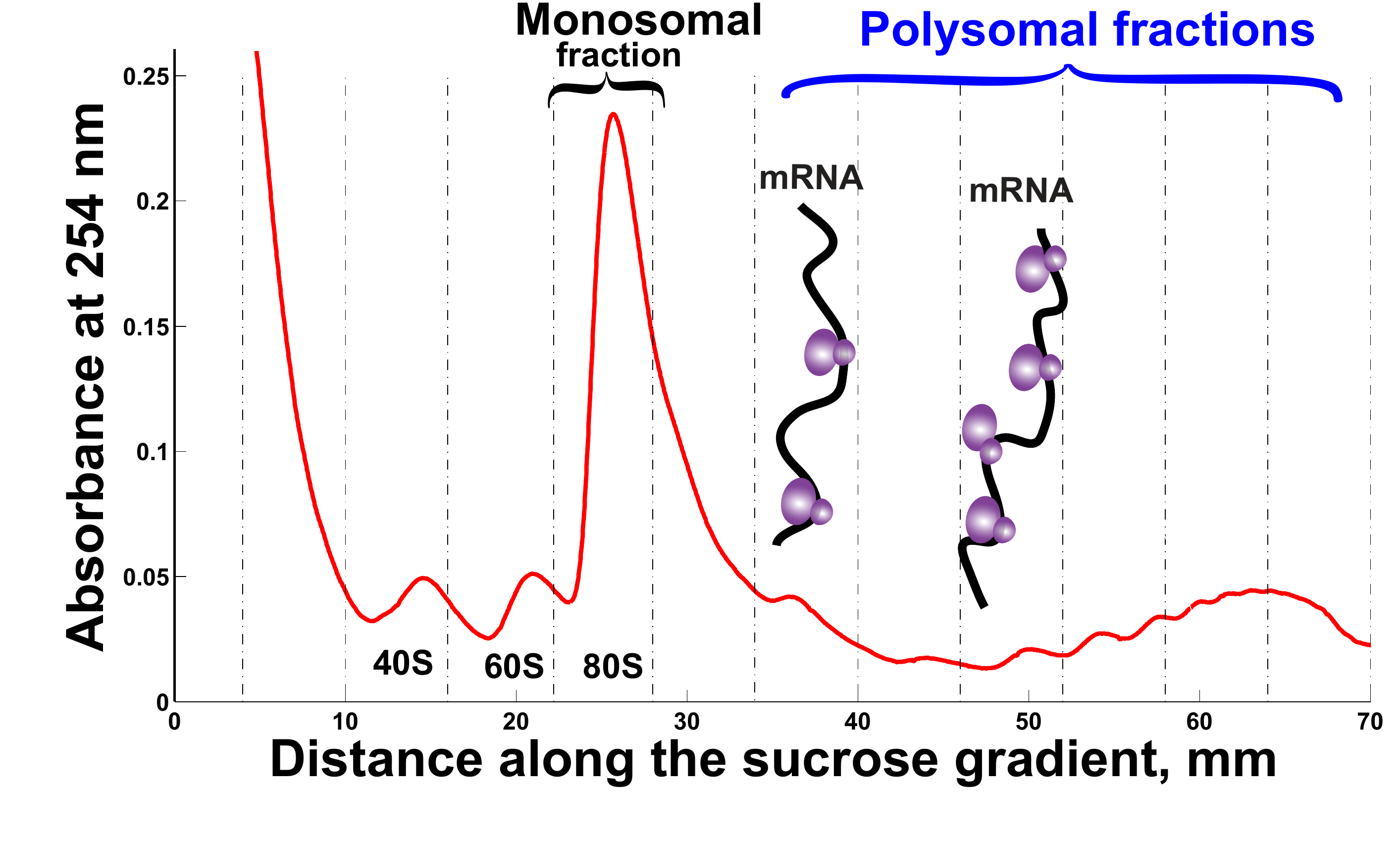}{A}  \\ [2em]  
	\inbL{Rev/Replicas_Mouse_2}{B}  \\ [2em] 
\end{minipage}
\hspace{0.1\textwidth} 
\begin{minipage}{0.45\textwidth} 
	\BoxPlots{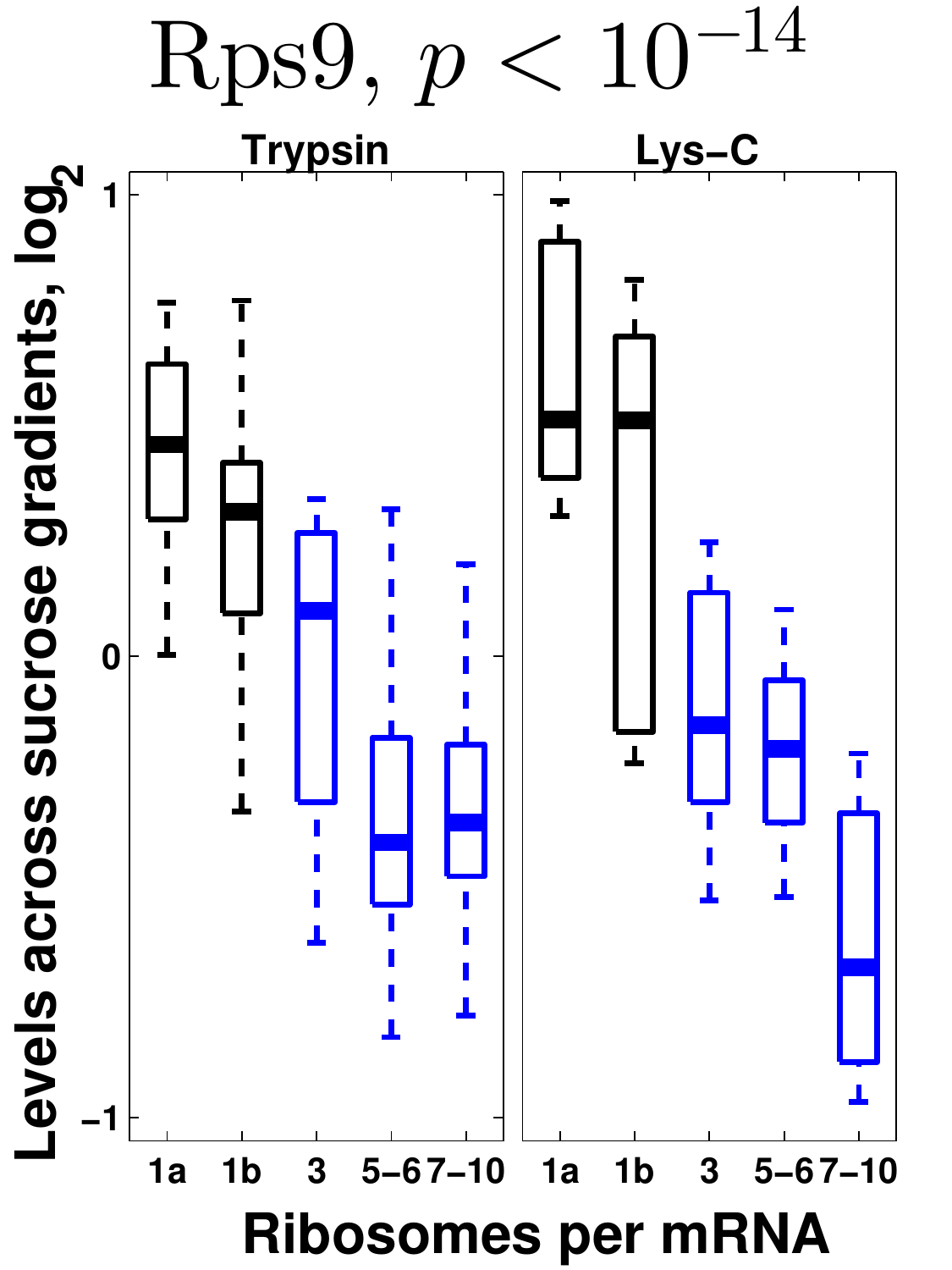}{C} 
	\includegraphics[width = .48\textwidth]{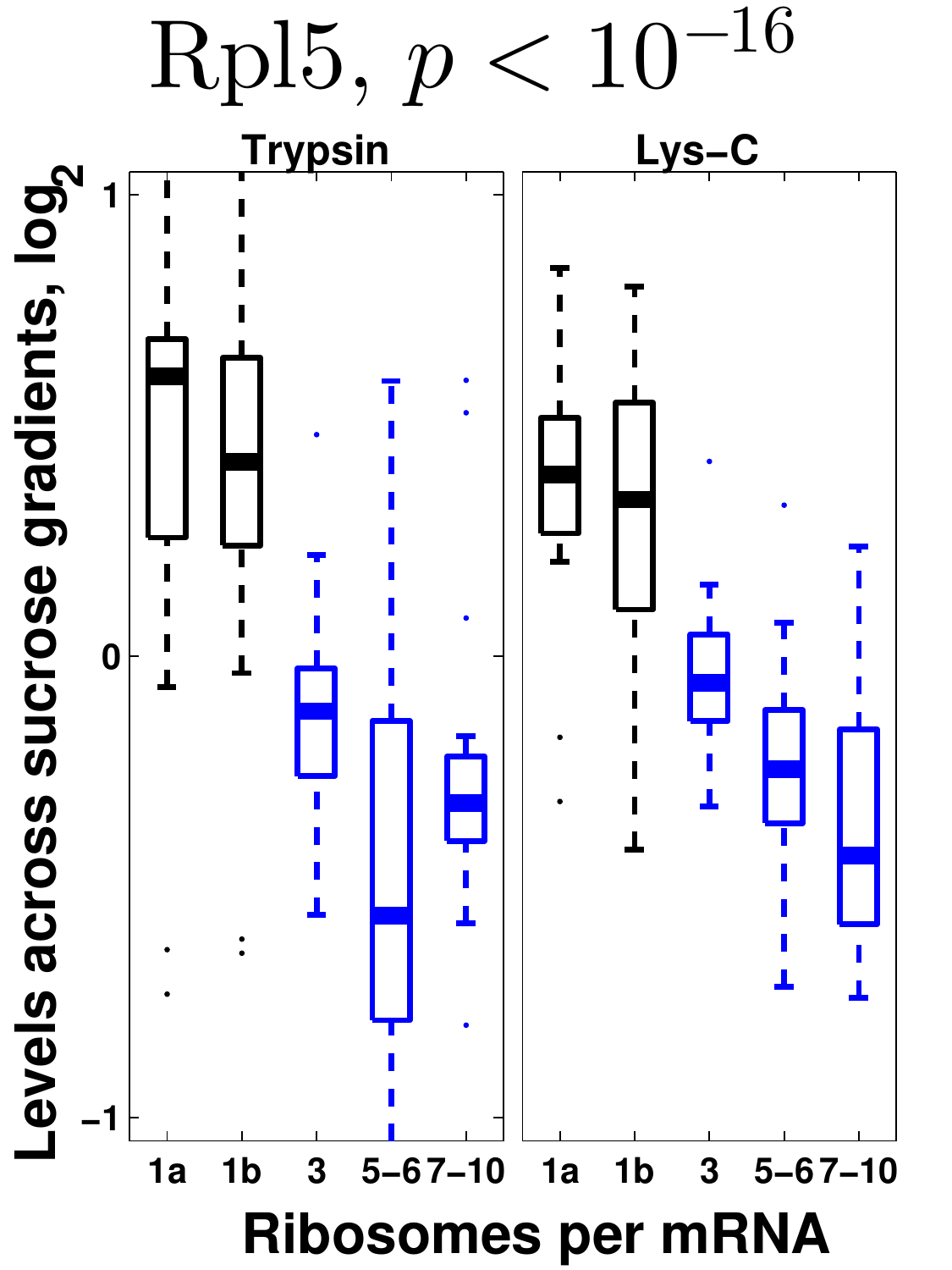} \\ [1em]
	\includegraphics[width = .48\textwidth]{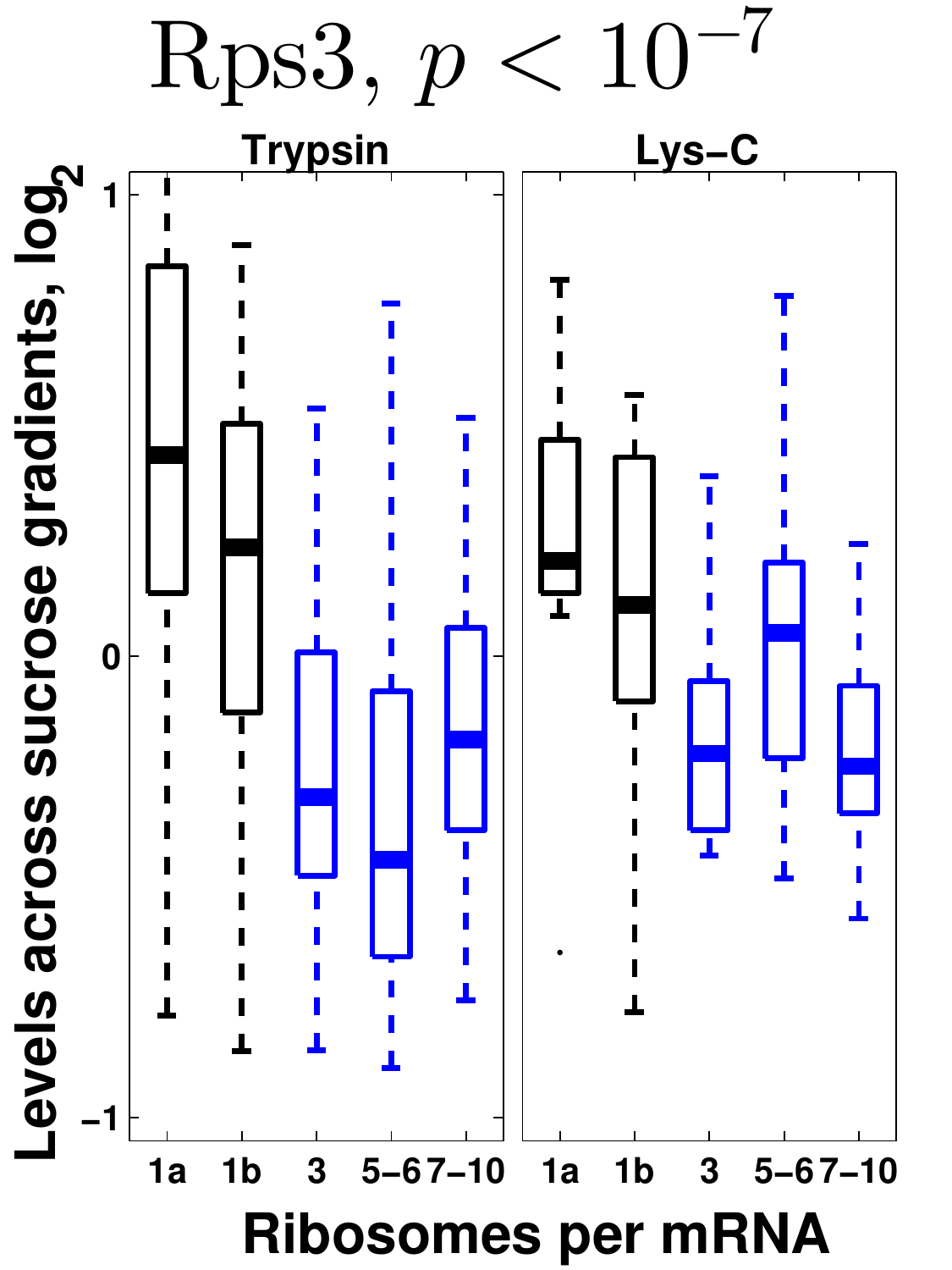} 
	\includegraphics[width = .48\textwidth]{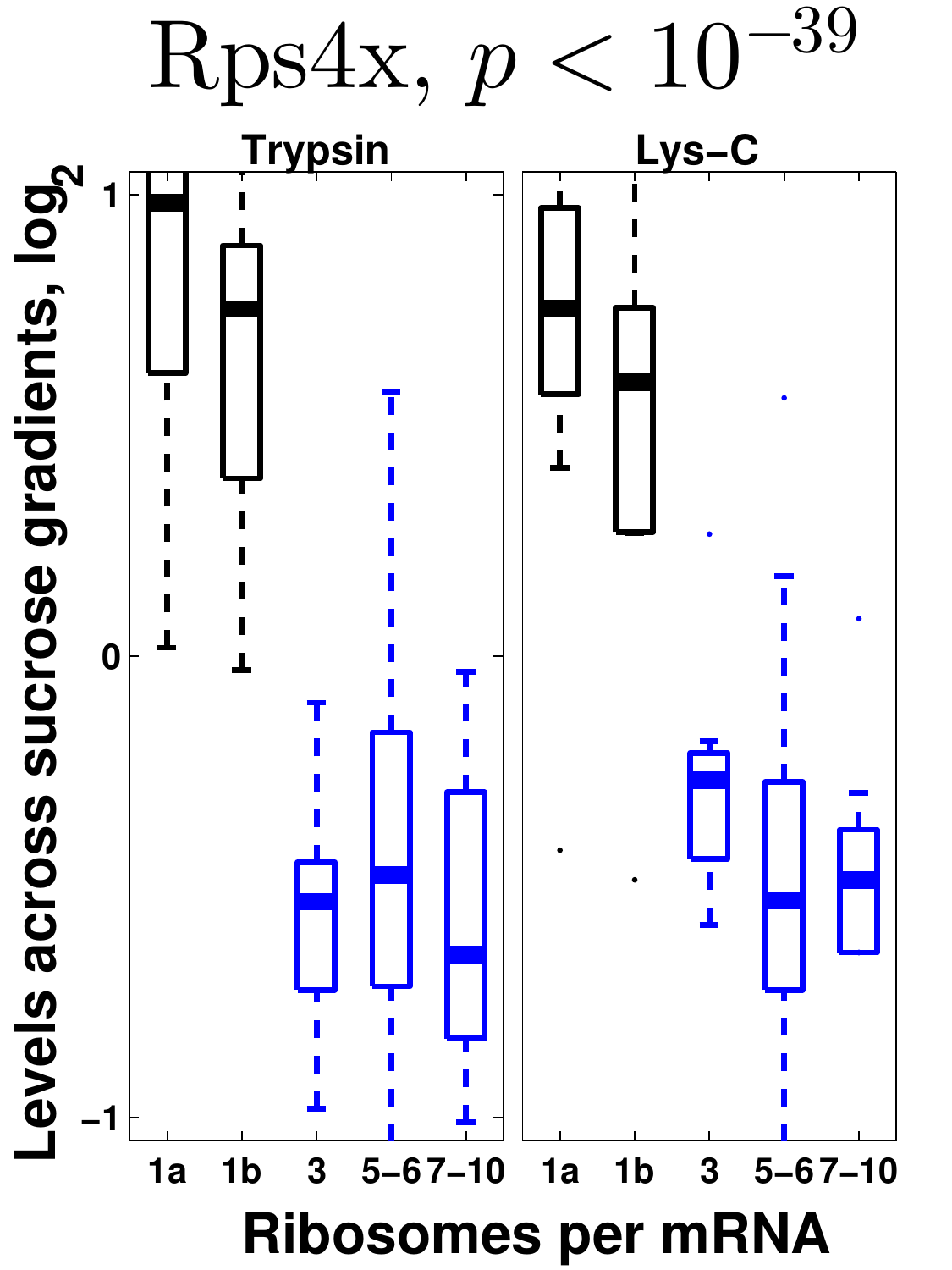} \\  
\end{minipage}
 \\ 
	\BoxPlotsS{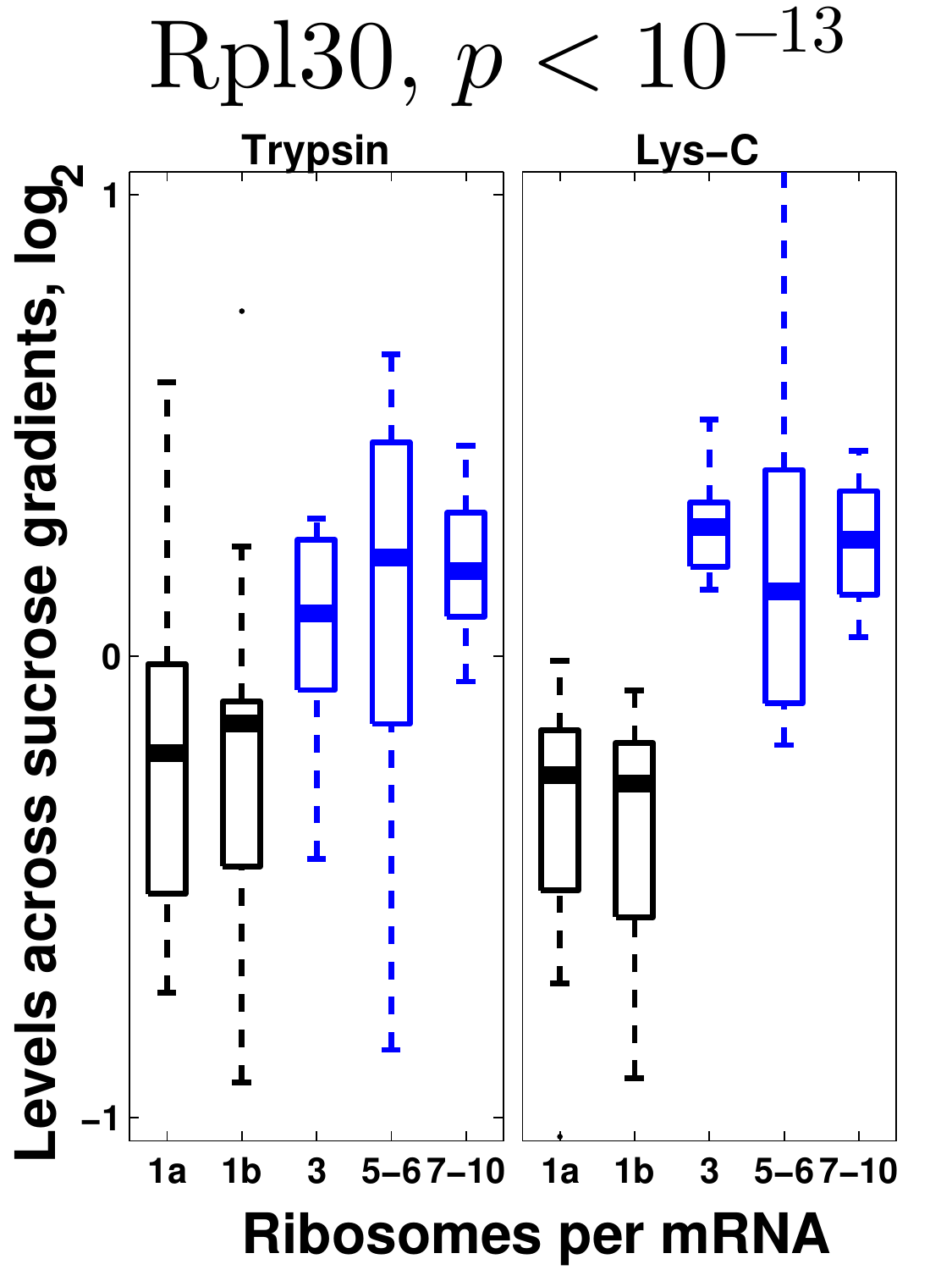}{D} 
	\includegraphics[width = .24\textwidth]{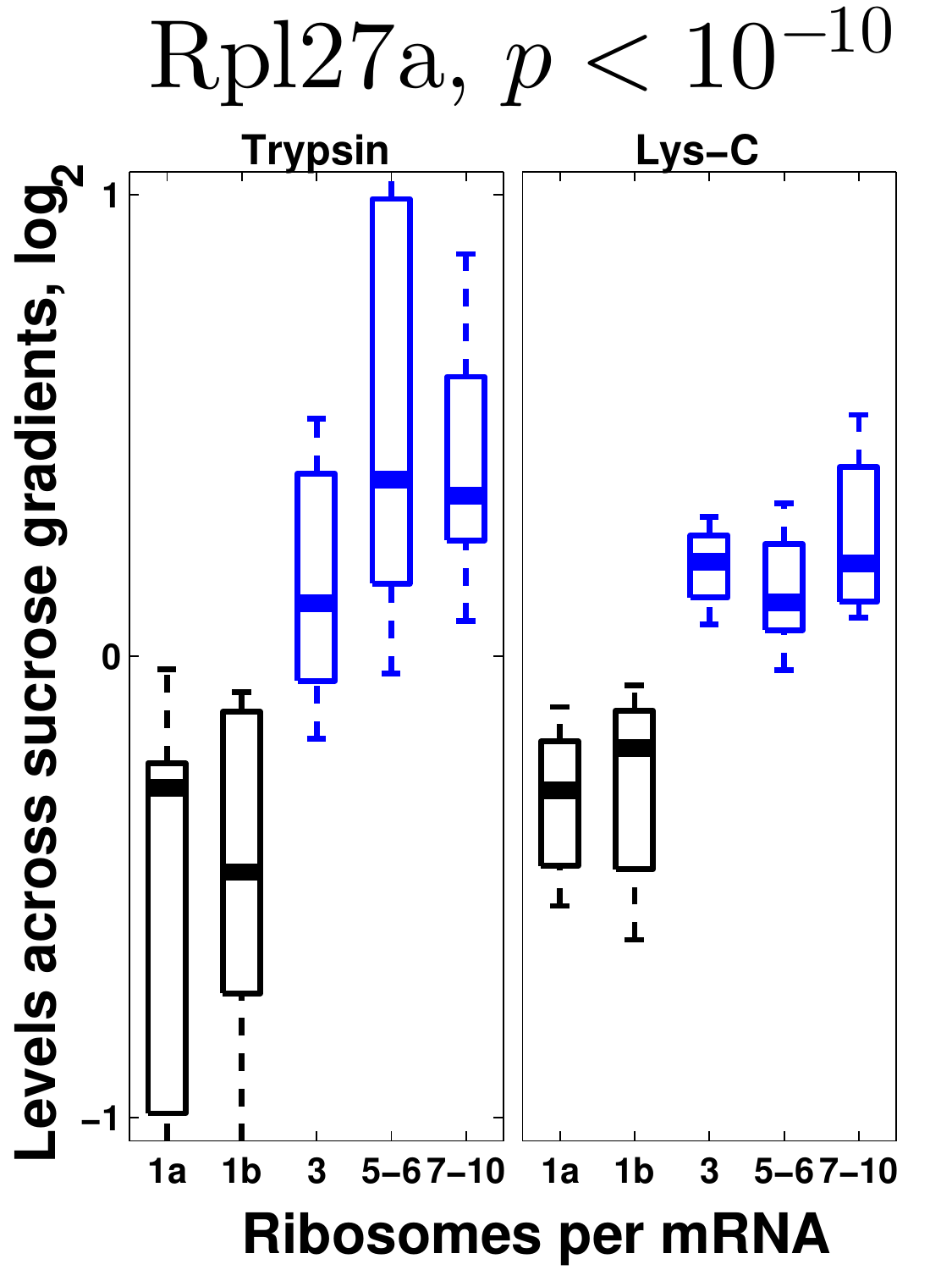}
	\includegraphics[width = .24\textwidth]{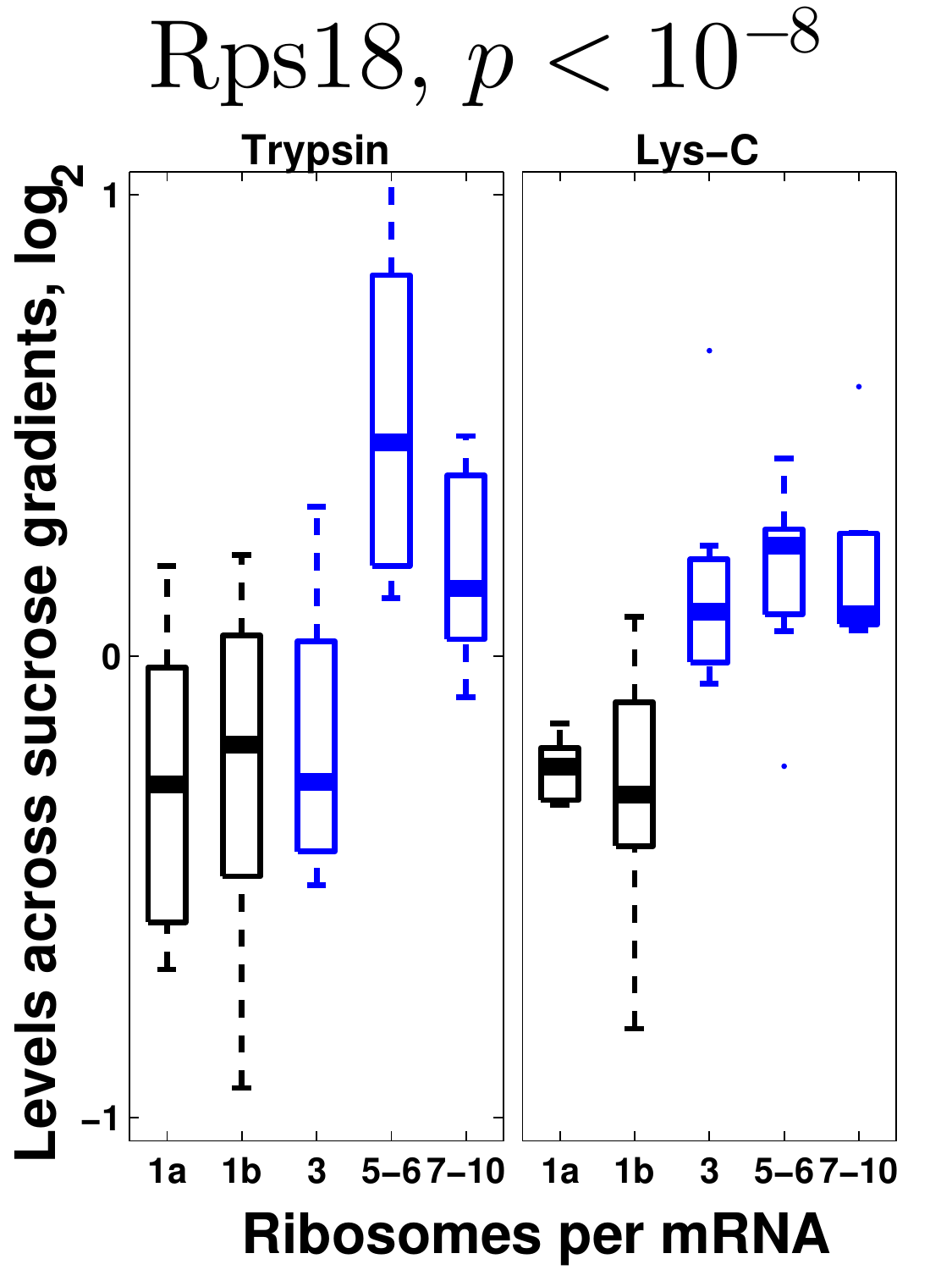} 
	\includegraphics[width = .24\textwidth]{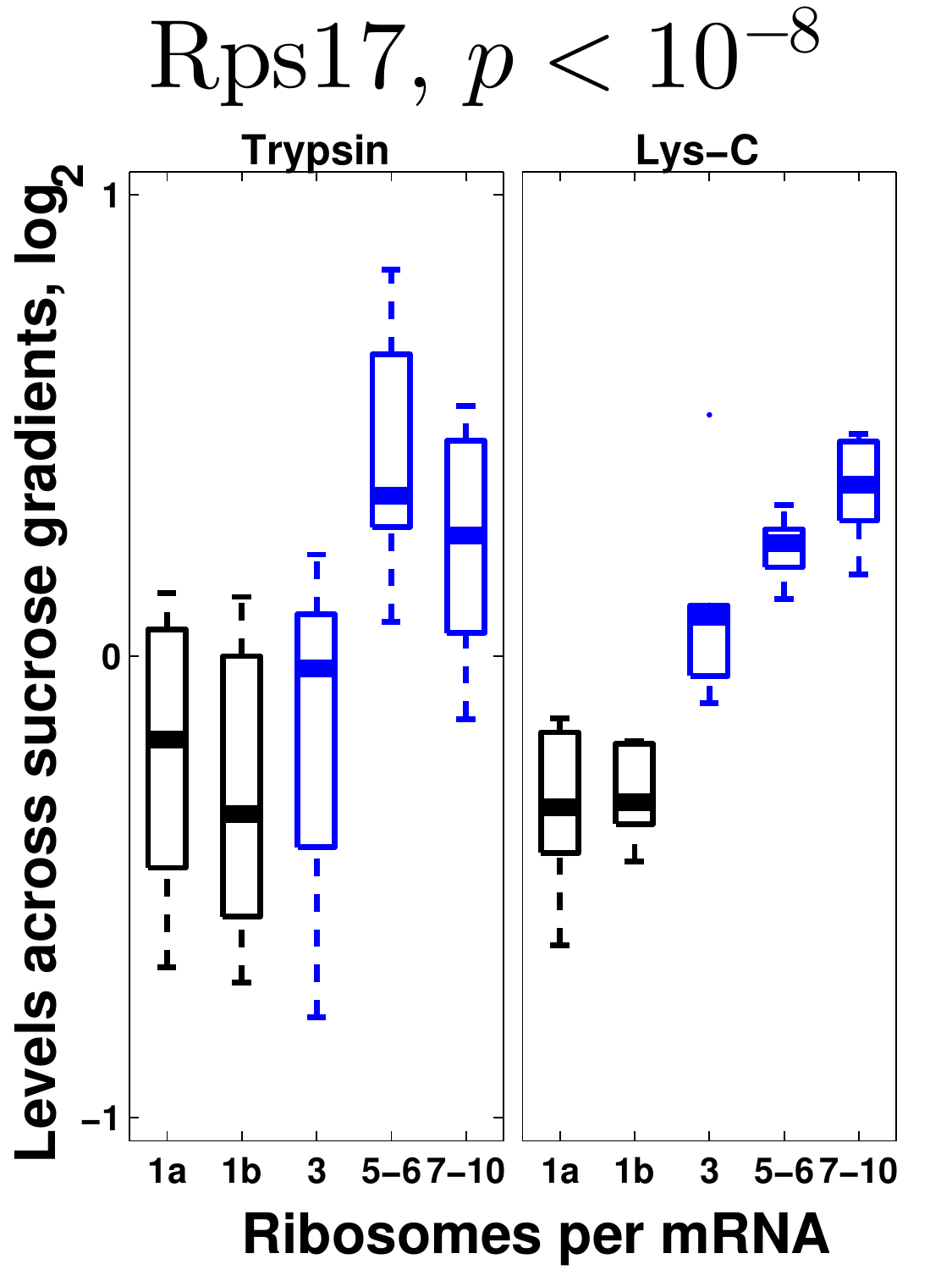}
\end{figure}
\newpage


\section*{Figure 2}
\vspace{8mm}
\begin{figure}[ht!]
	\includegraphics[width = .45\textwidth]{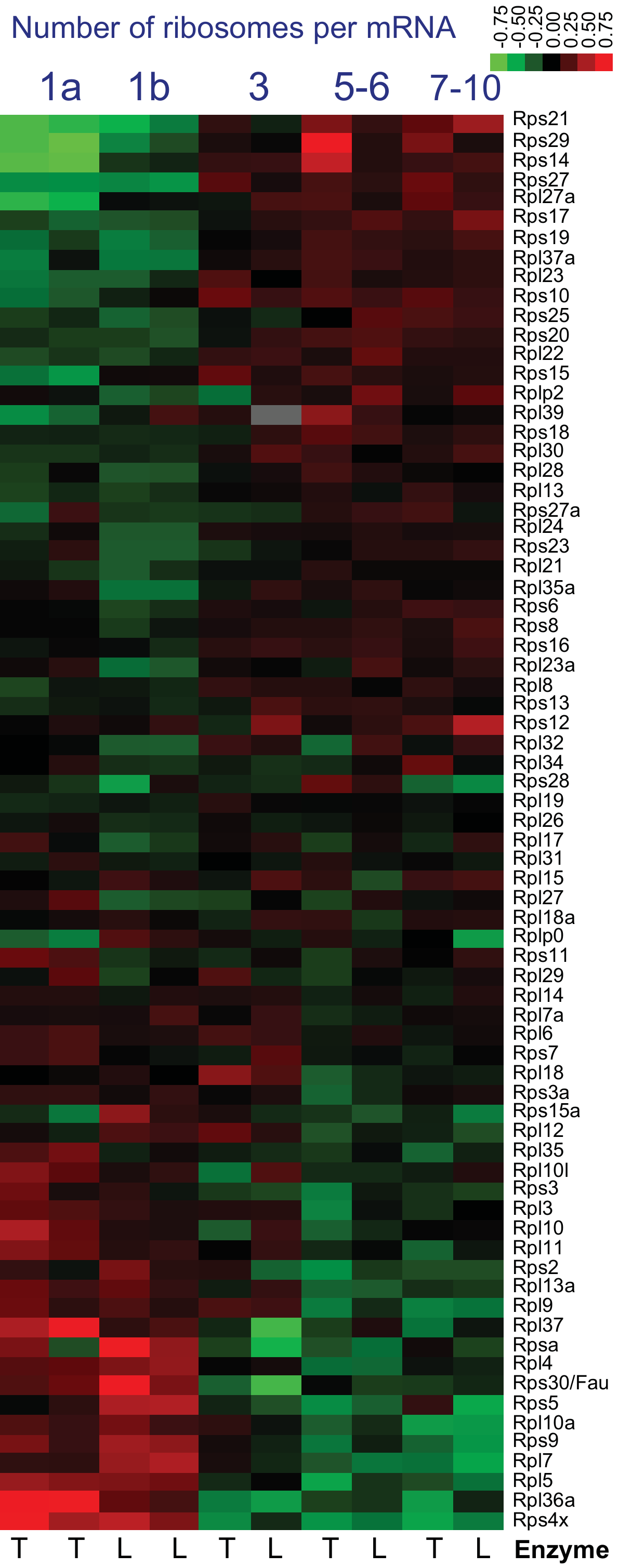}
\end{figure}
\newpage

\section*{Figure 3}
\vspace{8mm}
  \begin{minipage}{0.75\textwidth} 
 	\inaSm{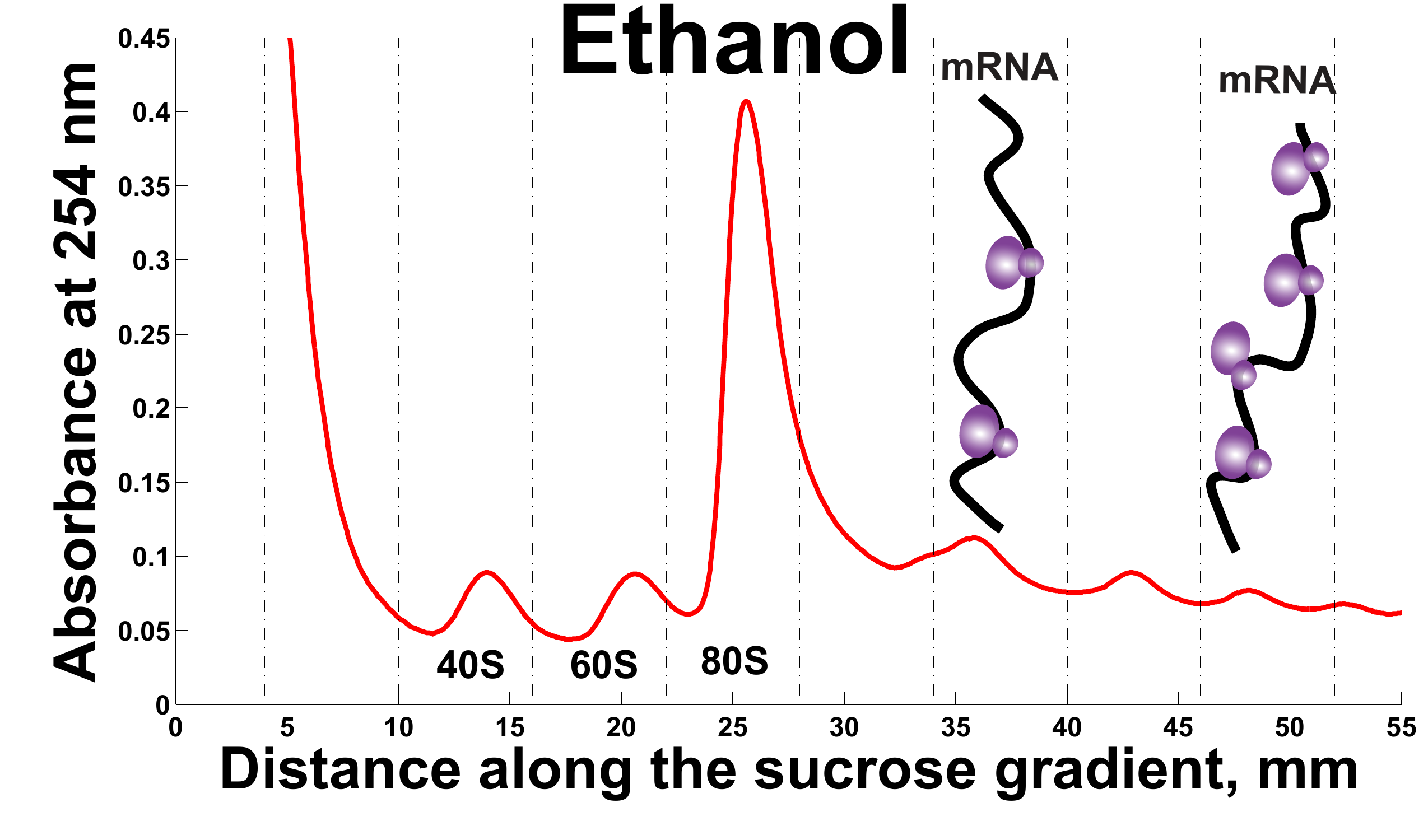}{A}
 	\inaSm{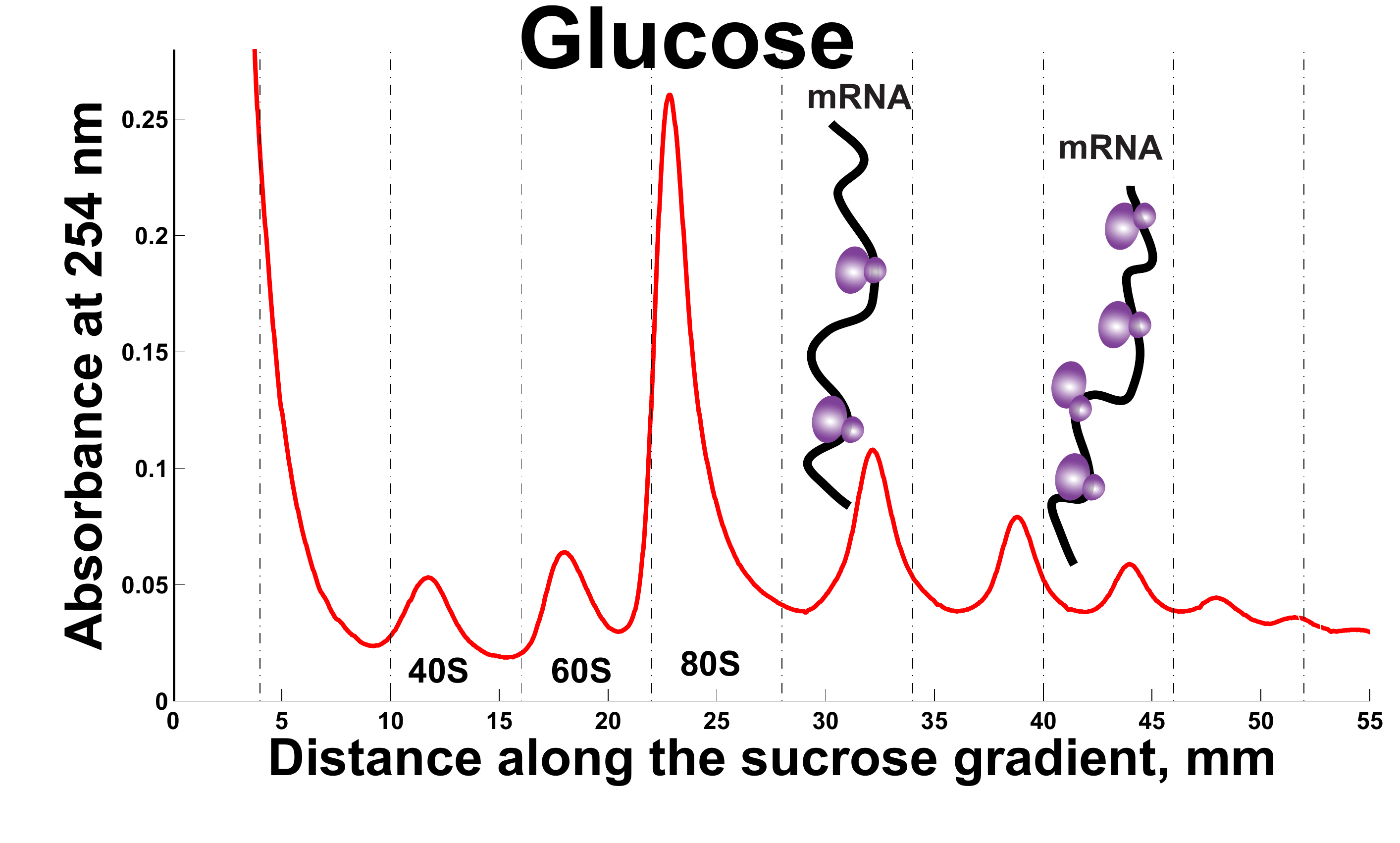}{B}  \\ [2em]
	\BoxPlotsM{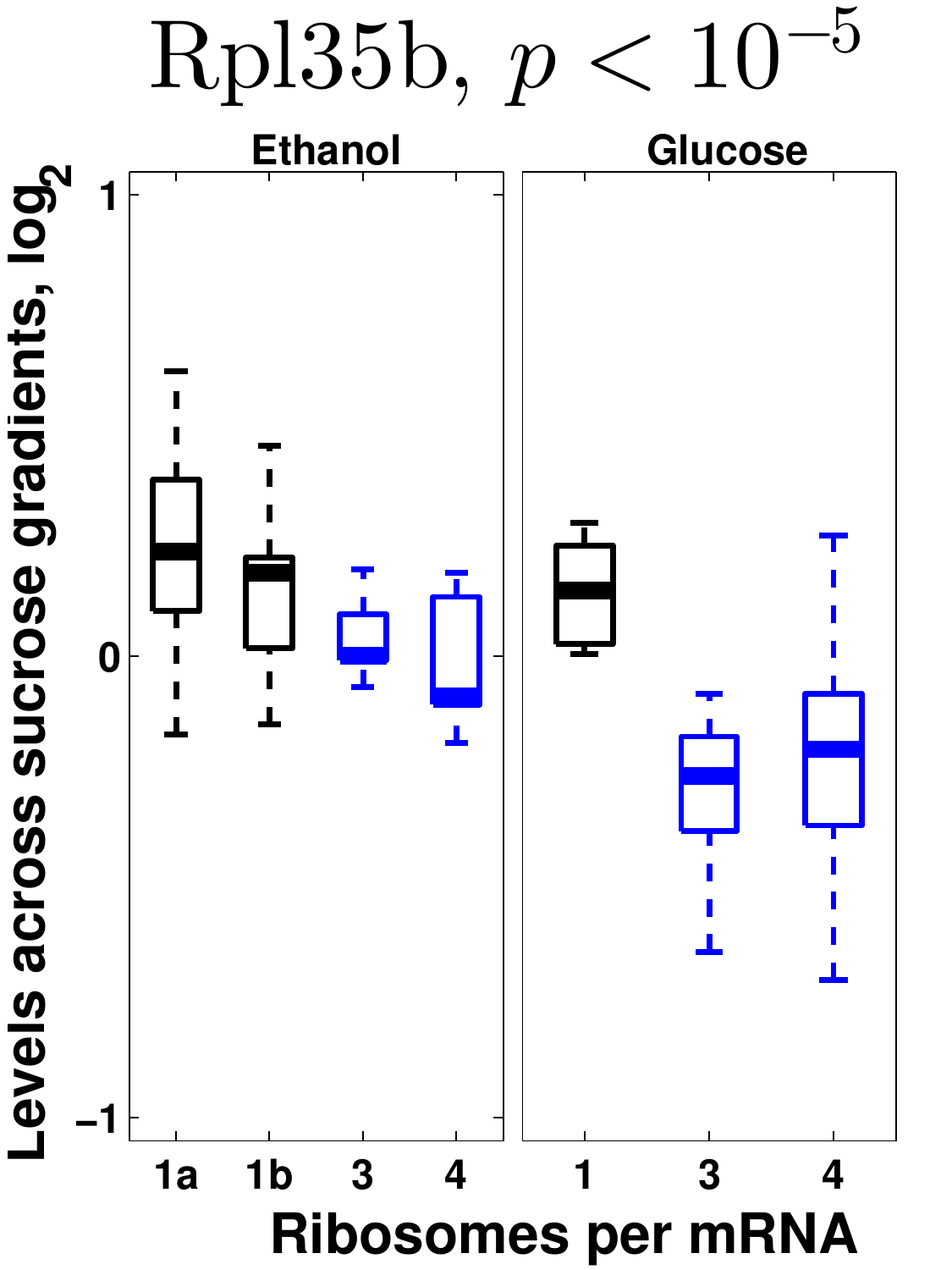}{C} \hspace{0.1\textwidth}
	\BoxPlotsM{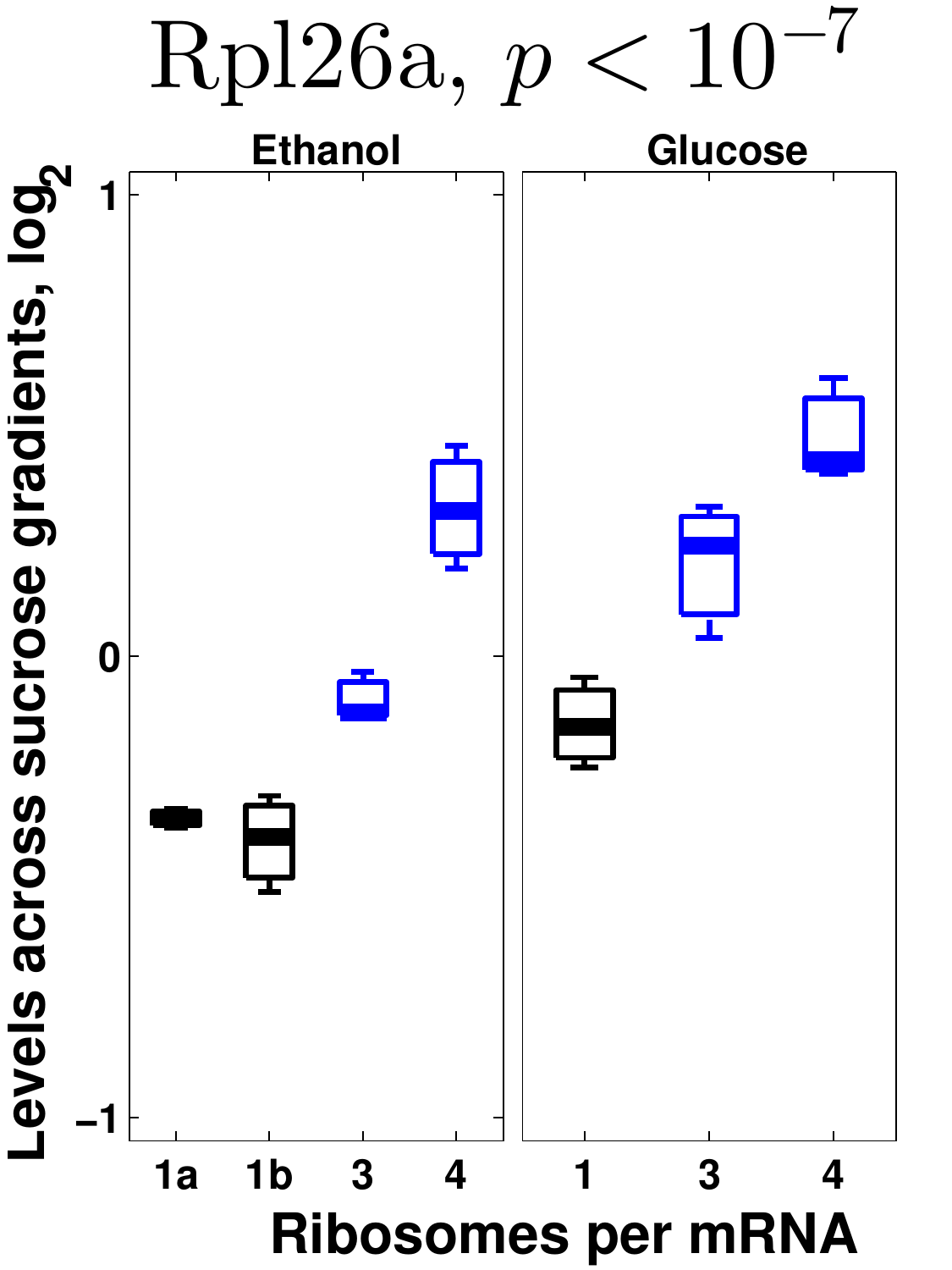}{D} \\ [2em]  
	\inbSm{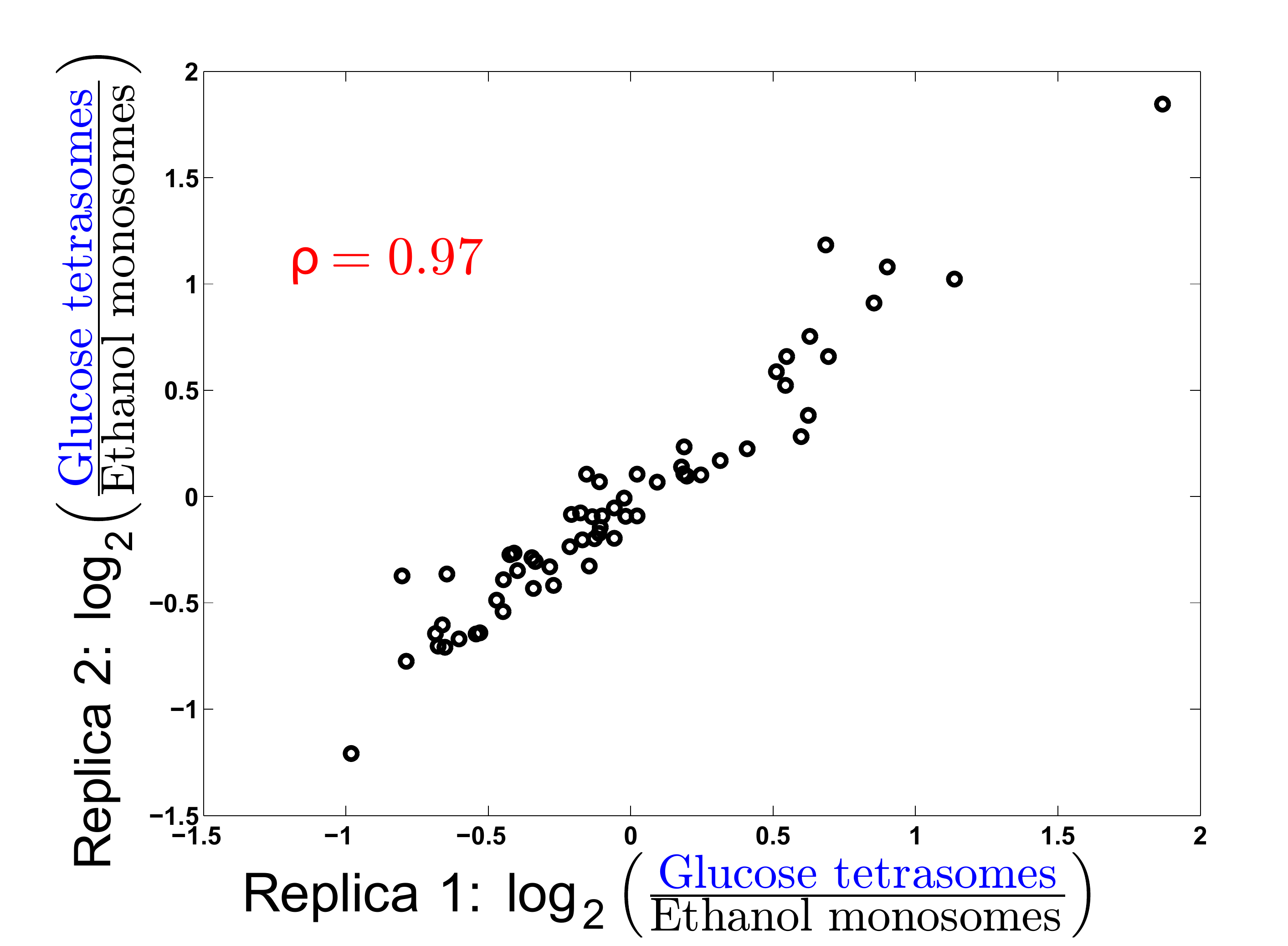}{F}   \hspace{0.05\textwidth}
	\inbSm{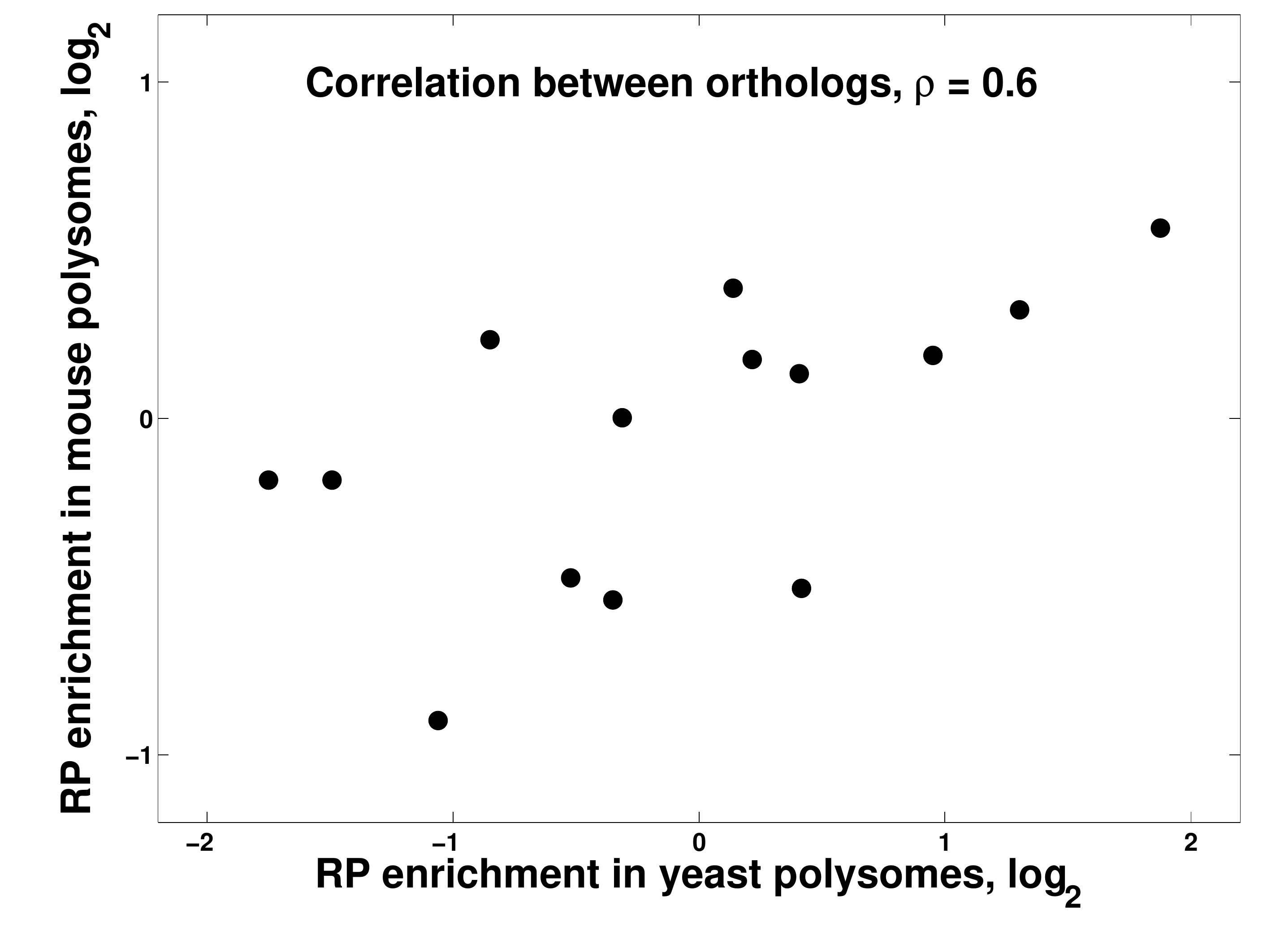}{G}  
  \end{minipage}
  \hspace{0.02\textwidth}
  \begin{minipage}{0.25\textwidth} 
	\inYeast{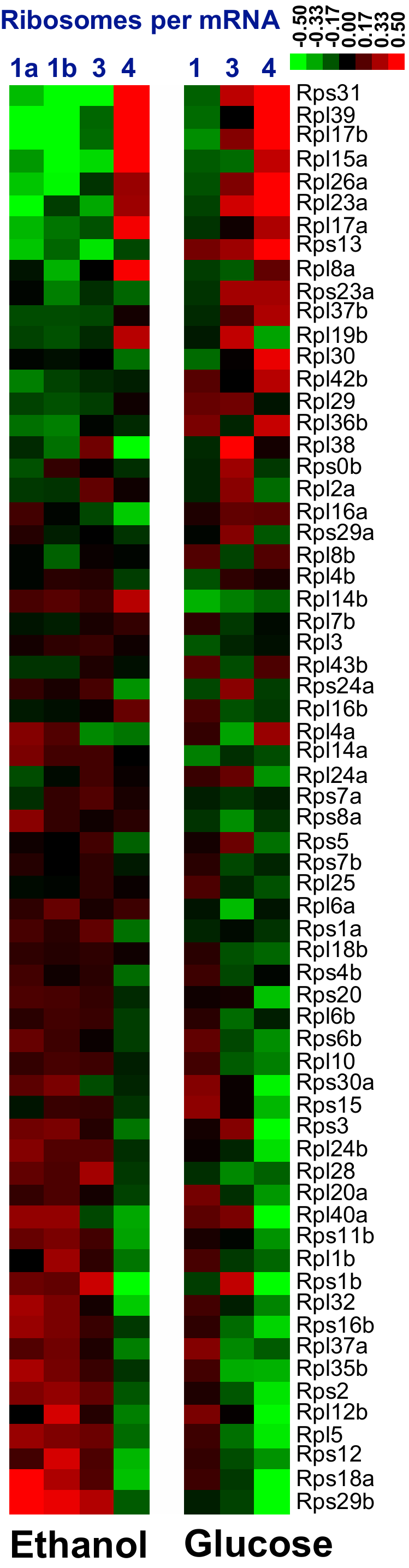}{E} \\ 
  \end{minipage}
\newpage

\section*{Figure 4}
\vspace{5mm}

\begin{figure}[ht!]
 \begin{tabular}{>{\centering\arraybackslash} p{.5\textwidth} 
 				 >{\centering\arraybackslash} p{.5\textwidth}}
 	  \large Yeast  &  \large   Mouse \& Human \\ [1em]
	\inb{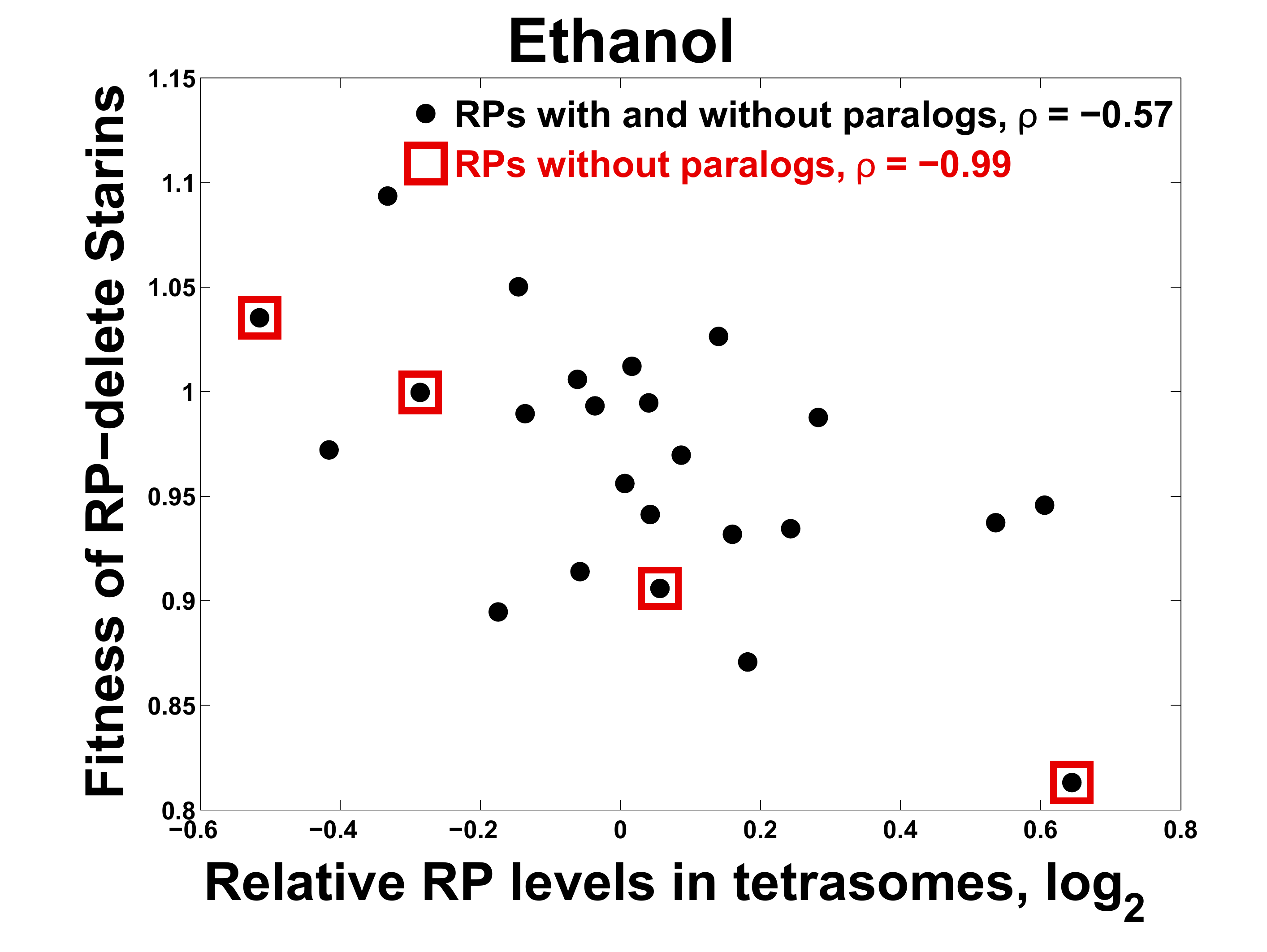}{A}   & 
		\inb{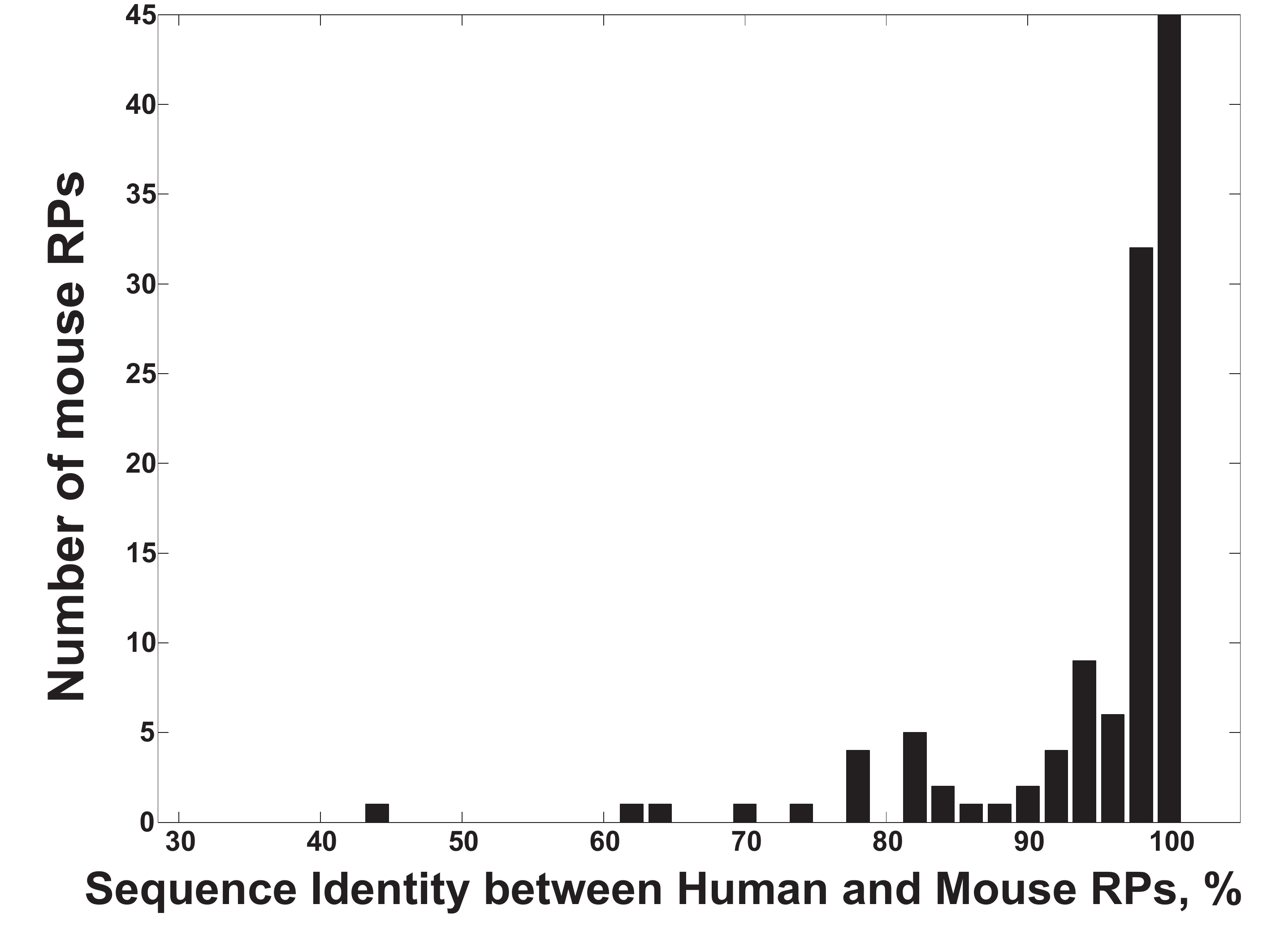}{D}  \\ [2em]
	\inb{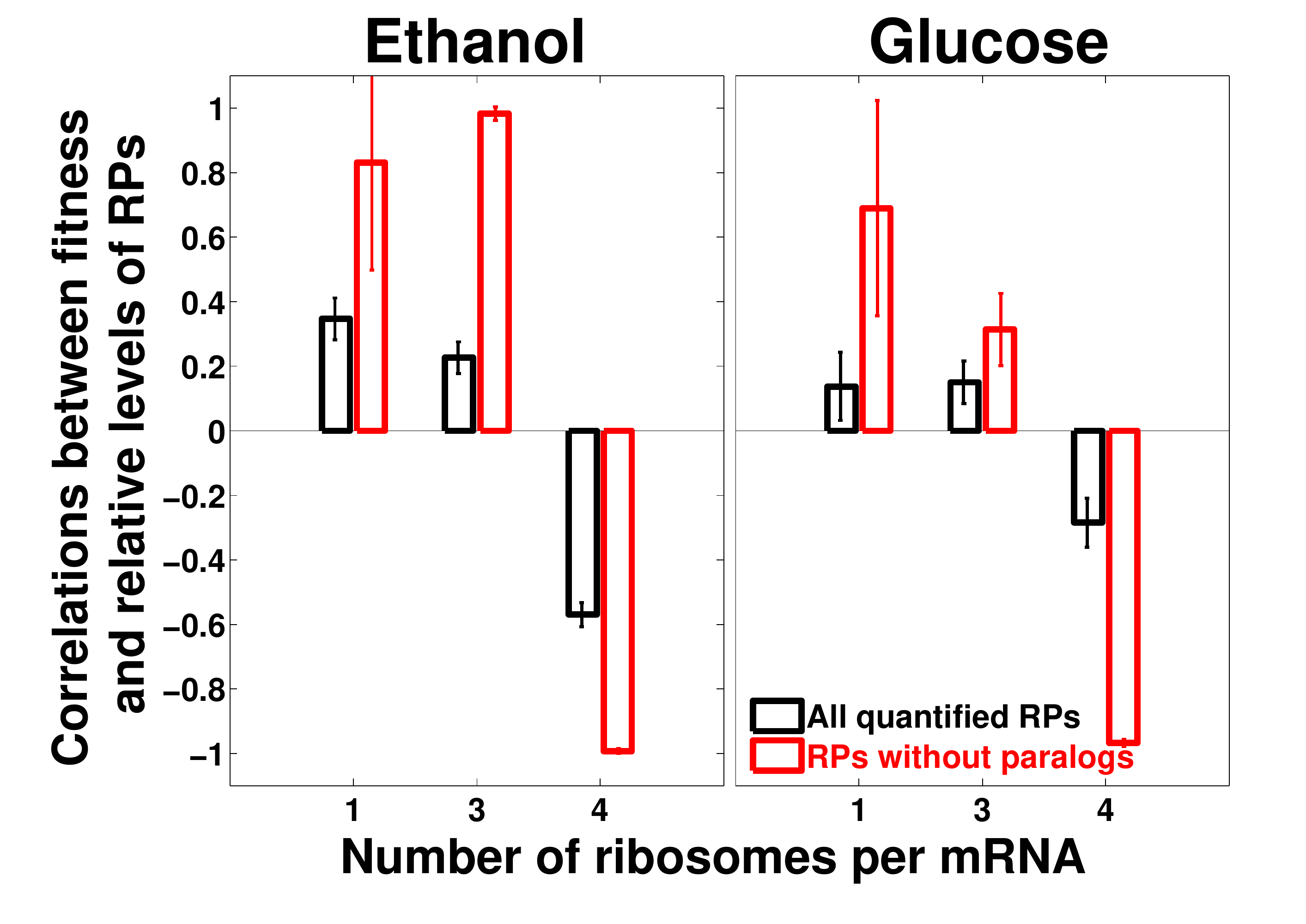}{B} &
		\inb{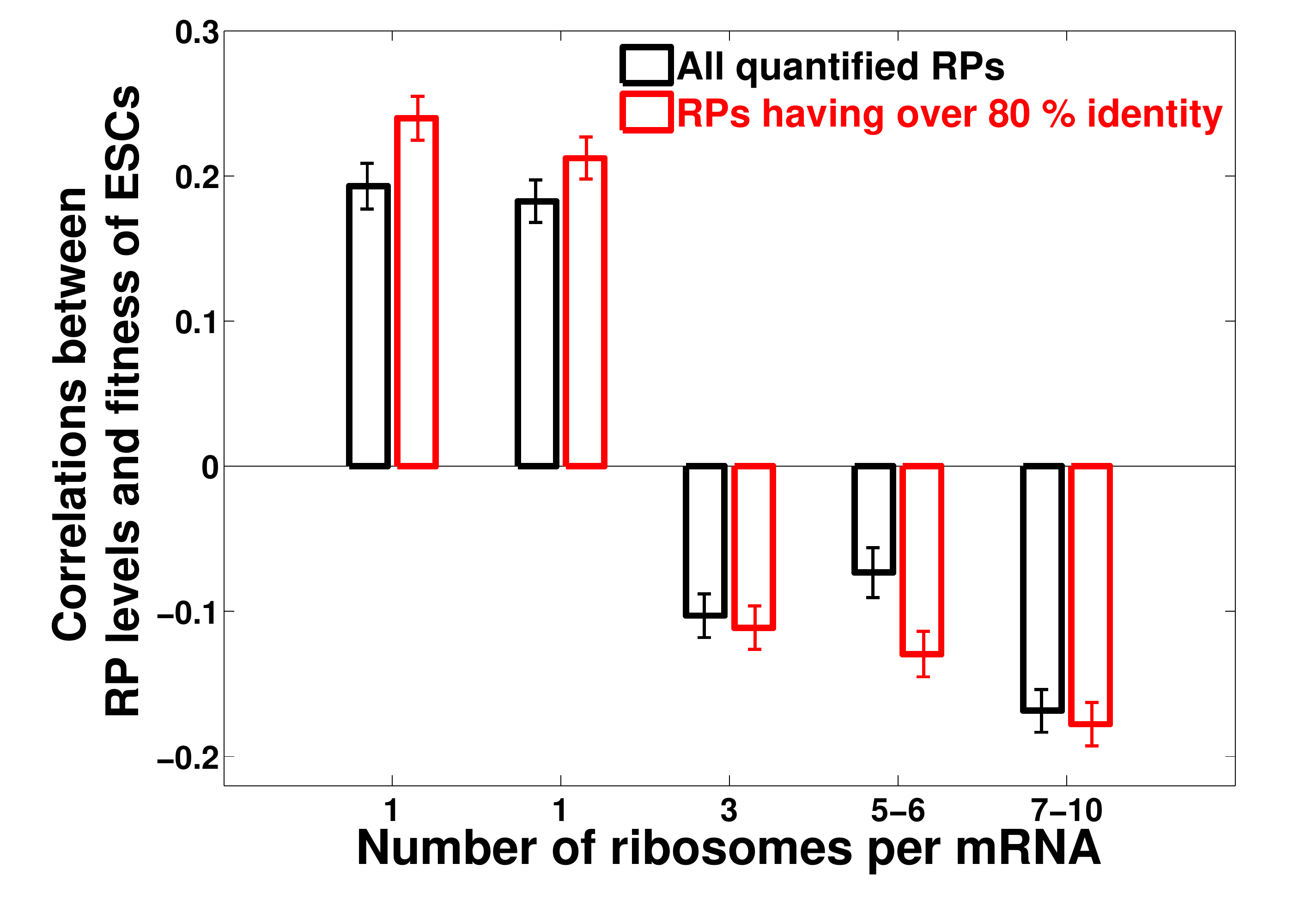}{E} \\[2em]
	\inb{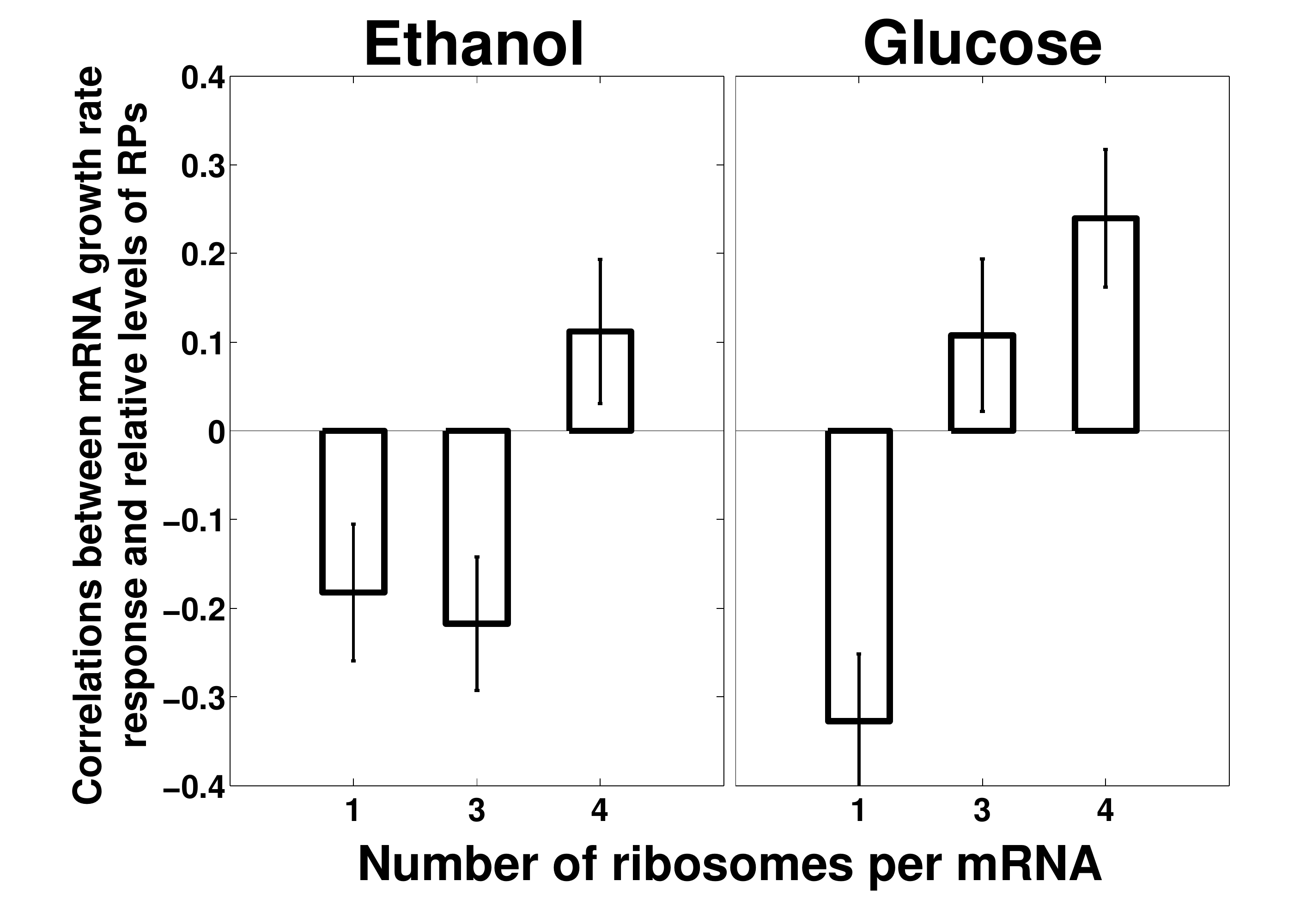}{C} &  \\
 \end{tabular}
\end{figure}
\clearpage

\section*{Figure E1}
\vspace{8mm}
\begin{figure}[ht!]
 \begin{tabular}{>{\arraybackslash} b{.12\textwidth} 
 				 >{\centering\arraybackslash} p{.45\textwidth}
 				 >{\centering\arraybackslash} p{.45\textwidth}}
 	& 	\Large Peptides per RP  & \Large Consistency \\ [1em]	 
 	\large {\bf Mouse} \emph{Trypsin} \newline \vspace{1cm} &
		\ina{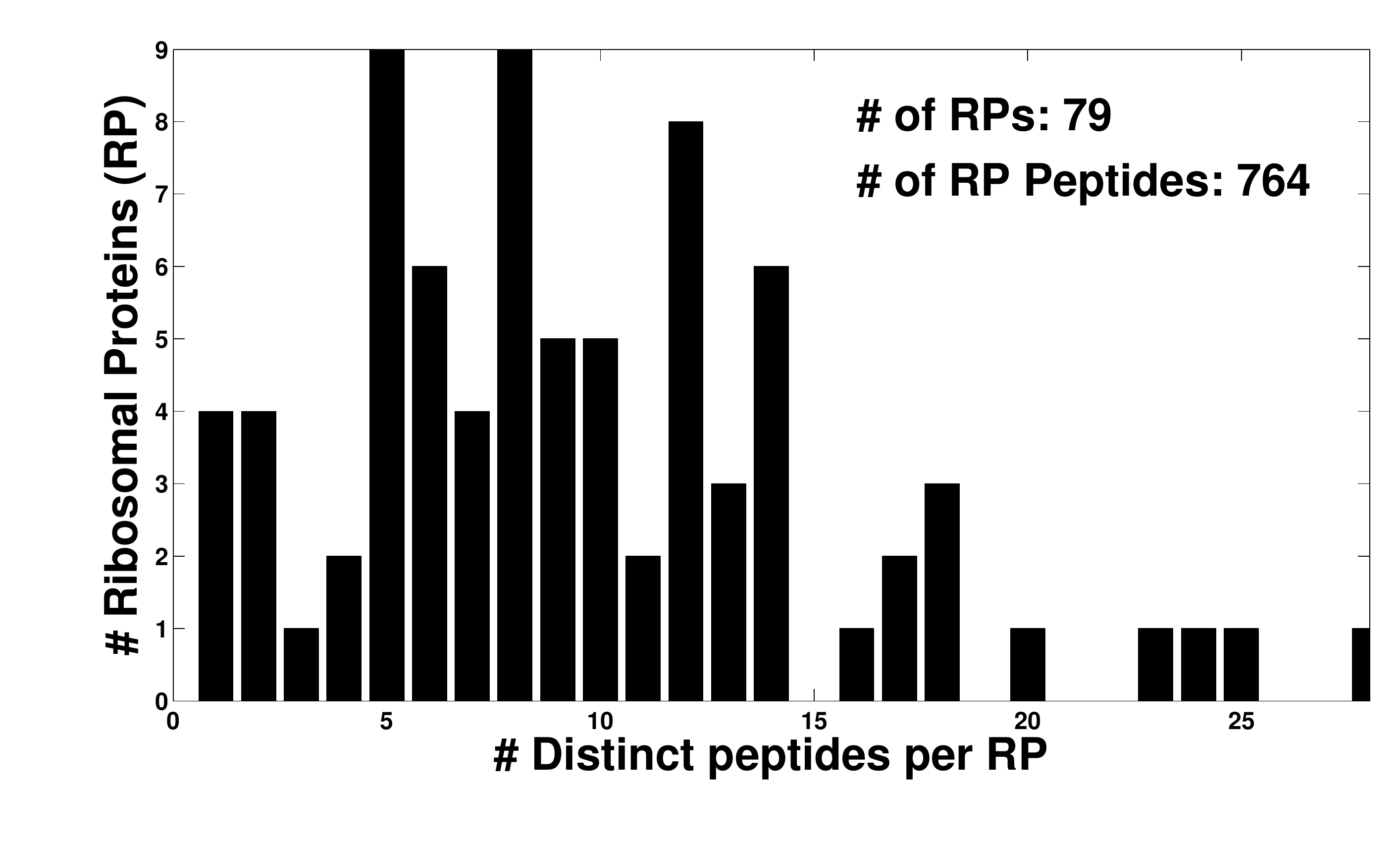}{A} & 
		\ina{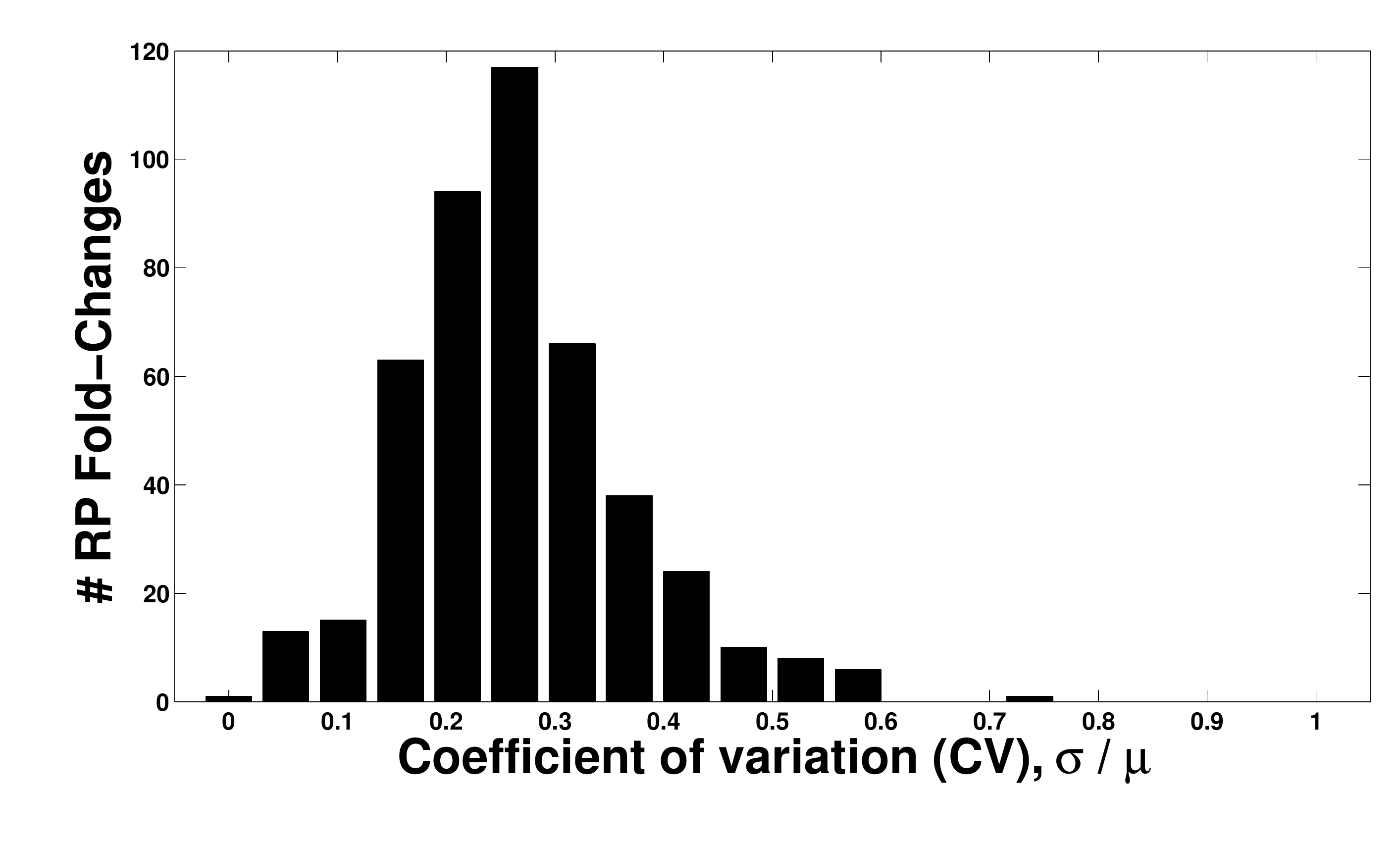}{B} \\ 		 	[2em] 
 	\large {\bf Mouse} \emph{Lys-C} \newline \vspace{1cm} &
		\ina{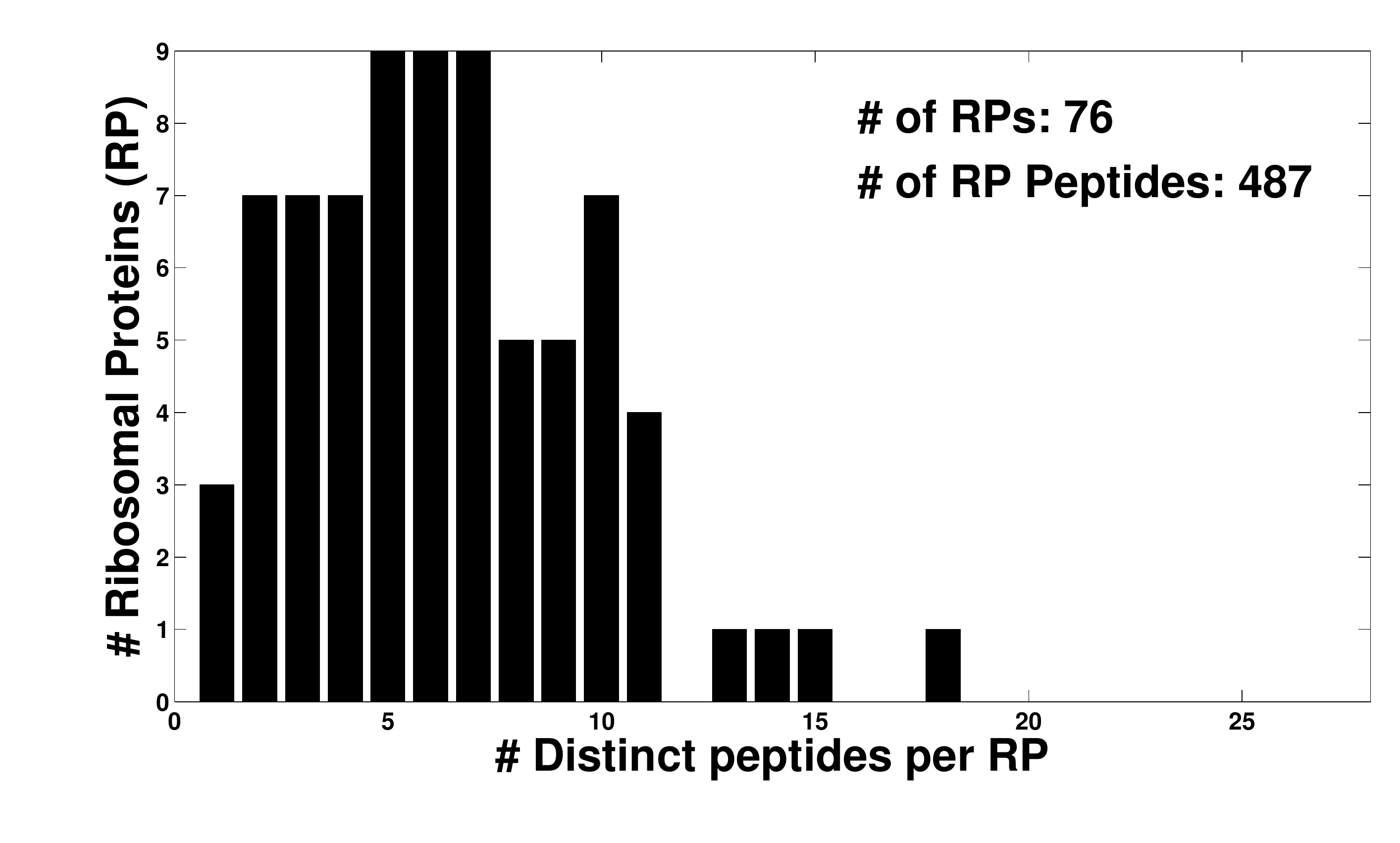}{C} & 
		\ina{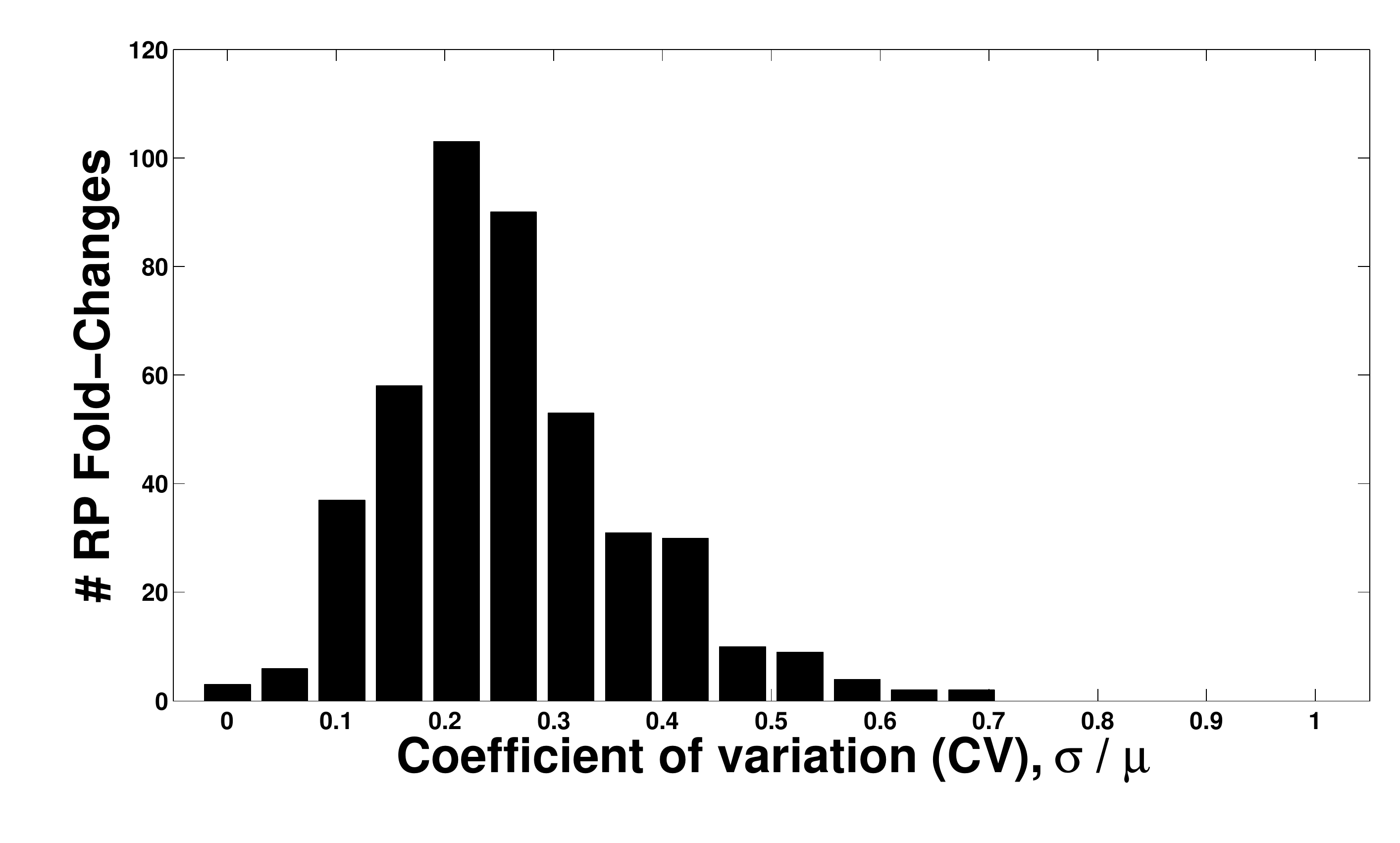}{D} \\ 			[2em] 	
 	\large {\bf Yeast}  \emph{Trypsin}   \newline \vspace{1cm} &
		\ina{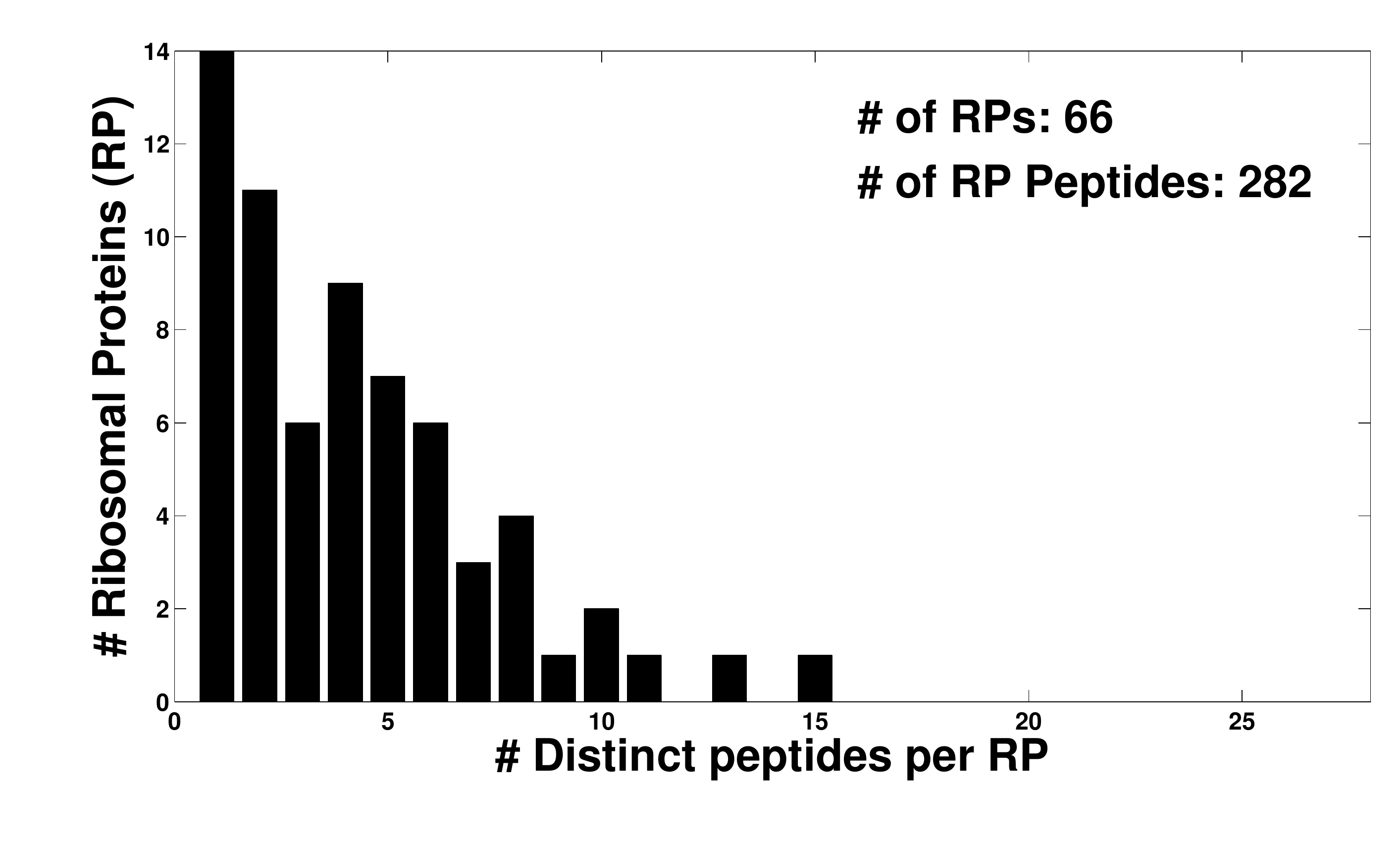}{E} & 
		\ina{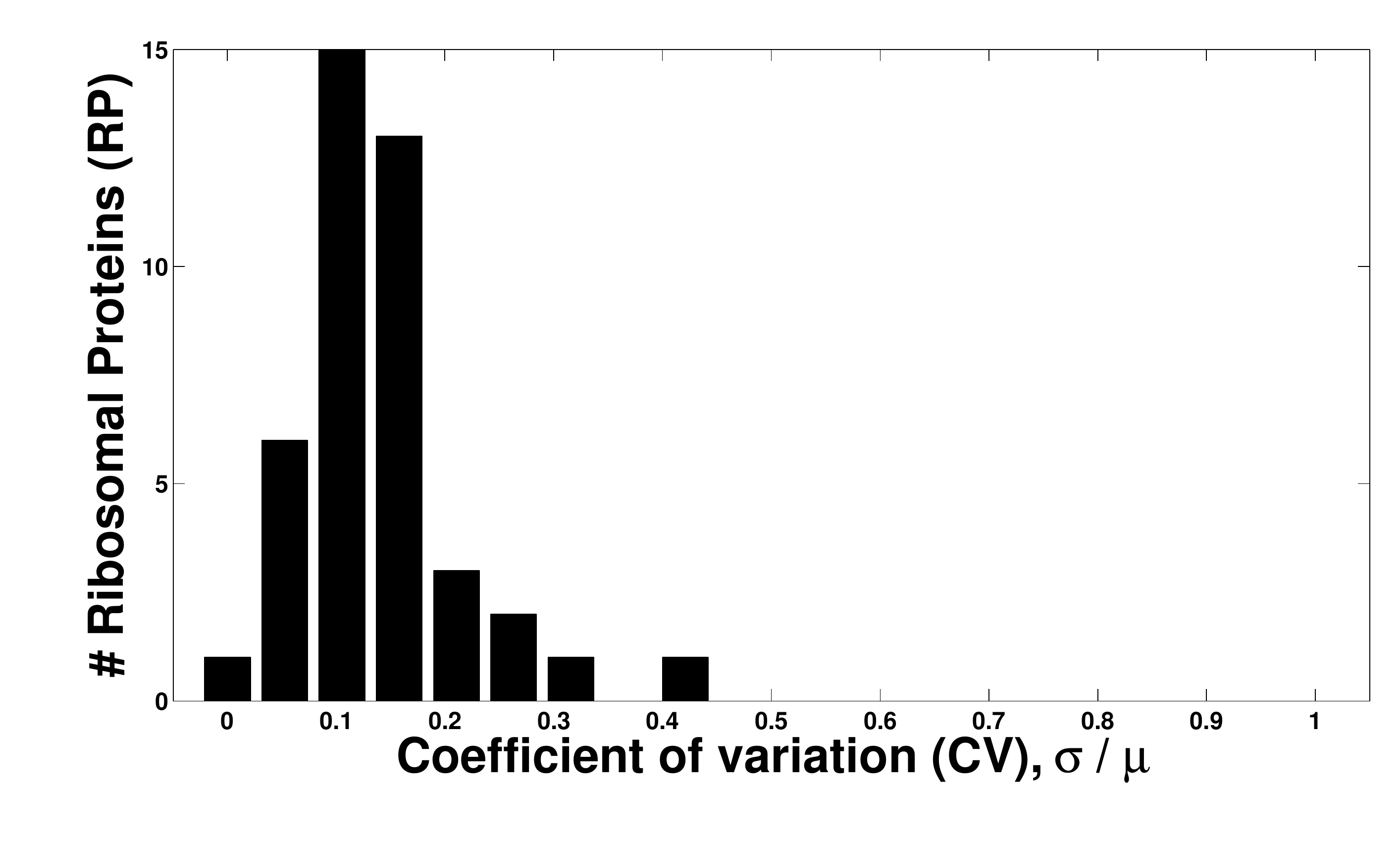}{F} \\ [1em] 			
 \end{tabular}
\end{figure}
\newpage


\section*{Figure E2}
\vspace{8mm}
\begin{figure}[ht!]
    \centering
 	\includegraphics[width=120 mm]{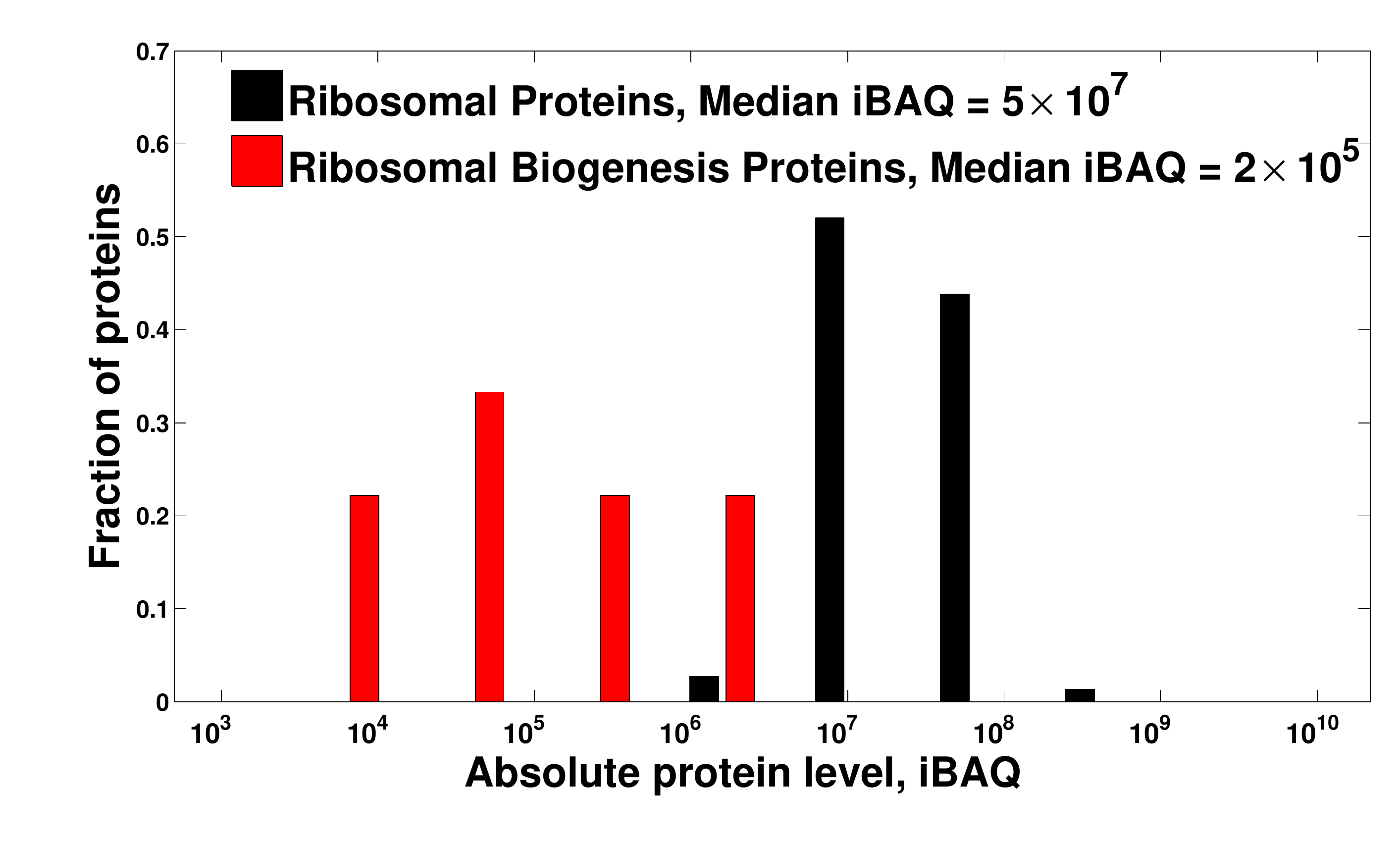}
\end{figure}
\newpage

\section*{Figure E3}
\vspace{8mm}
\begin{figure}[ht!]
 \begin{tabular}{>{\centering\arraybackslash} p{.38\textwidth} %
 				 >{\centering\arraybackslash} p{.62\textwidth}}
 	  \Large Mass--Spec  &  \Large Western Blots \\ [1.5em]
 	\IIms{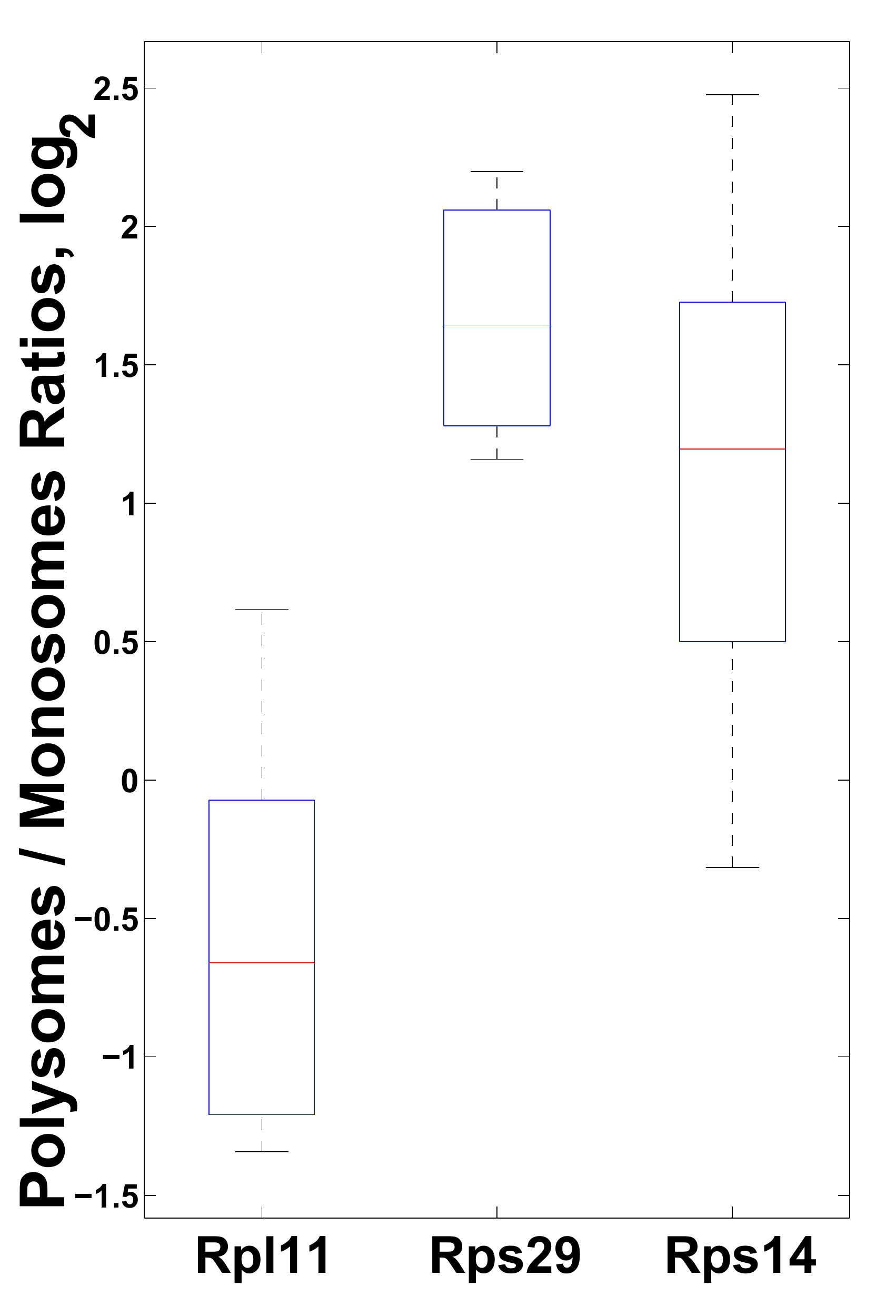}{A} &
 	\IIwb{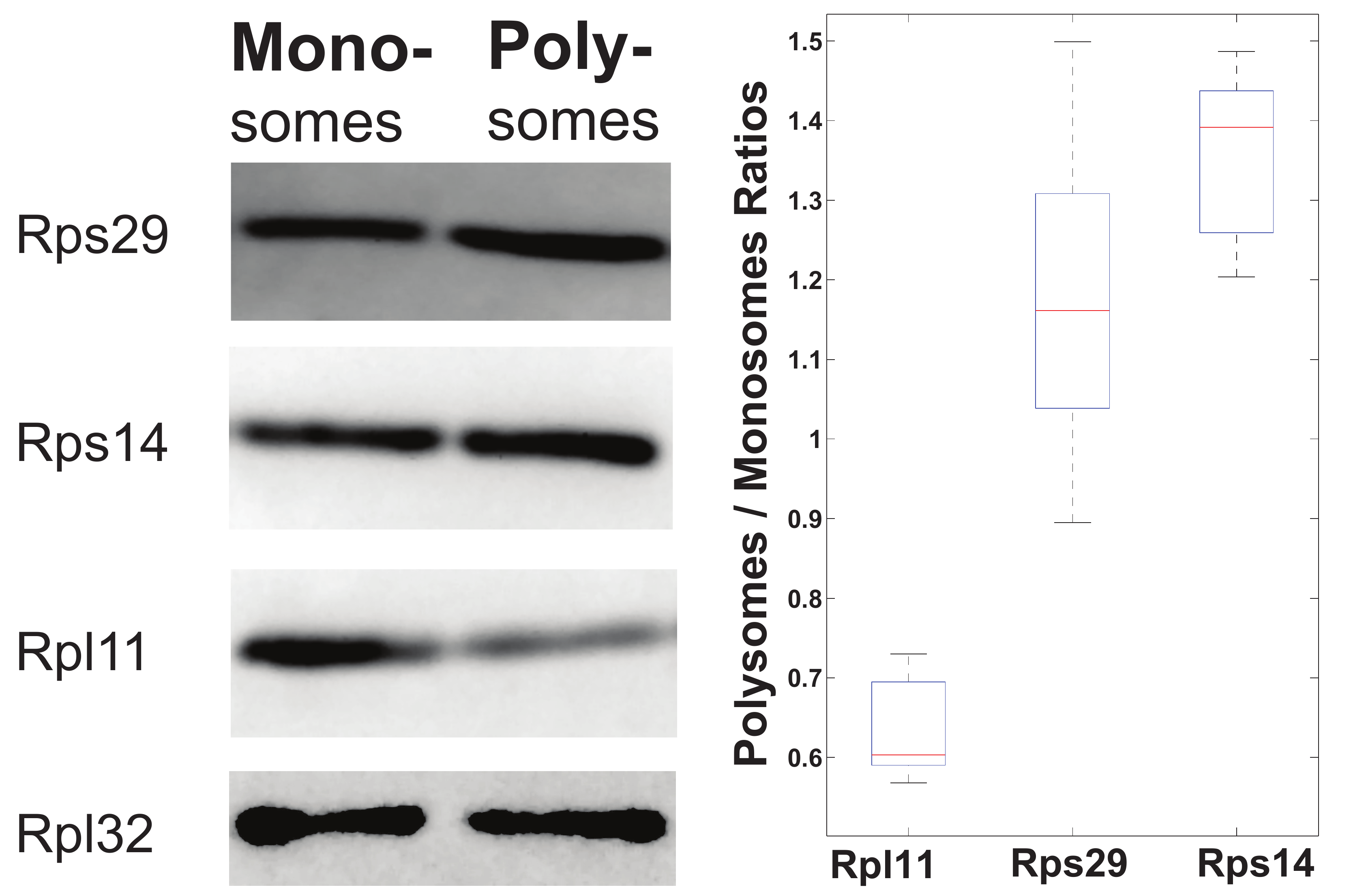}{B} \\
  \end{tabular}	
\end{figure}
\newpage

\section*{Figure E4}
\vspace{8mm}
\begin{figure}[ht!]
    \centering
 	\includegraphics[width = .9\textwidth]{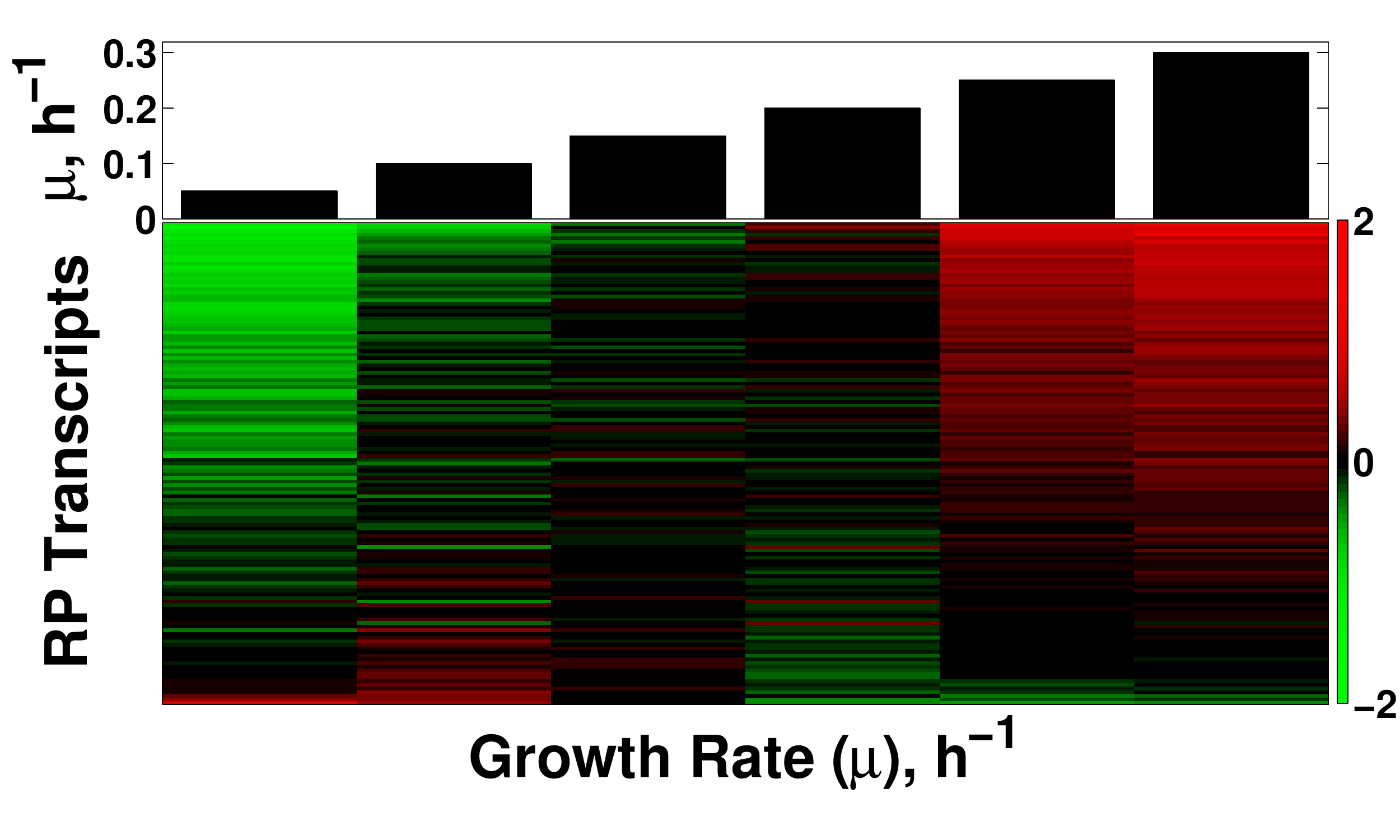}
\end{figure}
\newpage

\end{spacing}

\begin{thebibliography}{58}
\expandafter\ifx\csname natexlab\endcsname\relax\def\natexlab#1{#1}\fi

\bibitem[{Arava {\it et~al\/}(2003)Arava, Wang, Storey, Liu, Brown \&
  Herschlag}]{arava2003genome}
Arava Y, Wang Y, Storey JD, Liu CL, Brown PO, Herschlag D (2003) Genome-wide
  analysis of {mRNA} translation profiles in {{\it Saccharomyces cerevisiae}}.
  {\it Proceedings of the National Academy of Sciences\/} {\bf 100}: 3889--3894

\bibitem[{Ashe {\it et~al\/}(2000)Ashe, Susan \& Sachs}]{ashe2000glucose}
Ashe MP, Susan K, Sachs AB (2000) Glucose depletion rapidly inhibits
  translation initiation in yeast. {\it Molecular Biology of the Cell\/} {\bf
  11}: 833--848

\bibitem[{Ben-Shem {\it et~al\/}(2011)Ben-Shem, de~Loubresse, Melnikov, Jenner,
  Yusupova \& Yusupov}]{Xribo2011crystal_atomic}
Ben-Shem A, de~Loubresse NG, Melnikov S, Jenner L, Yusupova G, Yusupov M (2011)
  The structure of the eukaryotic ribosome at 3.0 {{\AA}} resolution. {\it
  Science\/} {\bf 334}: 1524--1529

\bibitem[{Ben-Shem {\it et~al\/}(2010)Ben-Shem, Jenner, Yusupova \&
  Yusupov}]{Xribo2010crystal}
Ben-Shem A, Jenner L, Yusupova G, Yusupov M (2010) Crystal structure of the
  eukaryotic ribosome. {\it Science\/} {\bf 330}: 1203--1209

\bibitem[{Bensimon {\it et~al\/}(2012)Bensimon, Heck \&
  Aebersold}]{Aebersold2012review}
Bensimon A, Heck AJ, Aebersold R (2012) {Mass spectrometry-based proteomics and
  network biology}. {\it Annual review of biochemistry\/} {\bf 81}: 379--405

\bibitem[{Bonven and Gull{\o}v(1979)Bonven \& Gull{\o}v}]{bonven1979peptide}
Bonven B, Gull{\o}v K (1979) Peptide chain elongation rate and ribosomal
  activity in {{\it Saccharomyces cerevisiae}} as a function of the growth
  rate. {\it Molecular and General Genetics MGG\/} {\bf 170}: 225--230

\bibitem[{Brauer {\it et~al\/}(2008)Brauer, Huttenhower, Airoldi, Rosenstein,
  Matese, Gresham, Boer, Troyanskaya \& Botstein}]{brauer_2008}
Brauer MJ, Huttenhower C, Airoldi EM, Rosenstein R, Matese JC, Gresham D, Boer
  VM, Troyanskaya OG, Botstein D (2008) Coordination of Growth Rate, Cell
  Cycle, Stress Response, and Metabolic Activity in Yeast. {\it Mol Biol
  Cell\/} {\bf 19}: 352--367

\bibitem[{Bui {\it et~al\/}(2013)Bui, von Appen, DiGuilio, Ori, Sparks,
  Mackmull, Bock, Hagen, Andr{\'e}s-Pons, Glavy {\it et~al\/}}]{npc_2013_beck}
Bui KH, von Appen A, DiGuilio AL, Ori A, Sparks L, Mackmull MT, Bock T, Hagen
  W, Andr{\'e}s-Pons A, Glavy JS, {\it et~al\/} (2013) Integrated structural
  analysis of the human nuclear pore complex scaffold. {\it Cell\/} {\bf 155}:
  1233--1243

\bibitem[{Castelli {\it et~al\/}(2011)Castelli, Lui, Campbell, Rowe, Zeef,
  Holmes, Hoyle, Bone, Selley, Sims {\it et~al\/}}]{castelli2011glucose}
Castelli LM, Lui J, Campbell SG, Rowe W, Zeef LA, Holmes LE, Hoyle NP, Bone J,
  Selley JN, Sims PF, {\it et~al\/} (2011) Glucose depletion inhibits
  translation initiation via {eIF4A} loss and subsequent {48S} preinitiation
  complex accumulation, while the pentose phosphate pathway is coordinately
  up-regulated. {\it Molecular biology of the cell\/} {\bf 22}: 3379--3393

\bibitem[{Castello {\it et~al\/}(2012)Castello, Fischer, Eichelbaum, Horos,
  Beckmann, Strein, Davey, Humphreys, Preiss, Steinmetz {\it
  et~al\/}}]{castello2012insights}
Castello A, Fischer B, Eichelbaum K, Horos R, Beckmann BM, Strein C, Davey NE,
  Humphreys DT, Preiss T, Steinmetz LM, {\it et~al\/} (2012) {Insights into RNA
  biology from an atlas of mammalian mRNA--binding proteins}. {\it Cell\/} {\bf
  149}: 1393--1406

\bibitem[{Dave {\it et~al\/}(2014)Dave, Granados-Principal, Zhu, Benz,
  Rabizadeh, Soon-Shiong, Yu, Shao, Li, Gilcrease {\it
  et~al\/}}]{dave2014targeting}
Dave B, Granados-Principal S, Zhu R, Benz S, Rabizadeh S, Soon-Shiong P, Yu KD,
  Shao Z, Li X, Gilcrease M, {\it et~al\/} (2014) {Targeting RPL39 and MLF2
  reduces tumor initiation and metastasis in breast cancer by inhibiting nitric
  oxide synthase signaling}. {\it Proceedings of the National Academy of
  Sciences\/} : 201320769

\bibitem[{De~Bortoli {\it et~al\/}(2006)De~Bortoli, Castellino, Lu, Deyo,
  Sturla, Adesina, Perlaky, Pomeroy, Lau, Man {\it
  et~al\/}}]{de2006medulloblastoma}
De~Bortoli M, Castellino RC, Lu XY, Deyo J, Sturla LM, Adesina AM, Perlaky L,
  Pomeroy SL, Lau CC, Man TK, {\it et~al\/} (2006) Medulloblastoma outcome is
  adversely associated with overexpression of {EEF1D, RPL30, and RPS20} on the
  long arm of chromosome 8. {\it BMC cancer\/} {\bf 6}: 223

\bibitem[{De~Keersmaecker {\it et~al\/}(2013)De~Keersmaecker, Atak, Li,
  Vicente, Patchett, Girardi, Gianfelici, Geerdens, Clappier, Porcu {\it
  et~al\/}}]{de2013exome}
De~Keersmaecker K, Atak ZK, Li N, Vicente C, Patchett S, Girardi T, Gianfelici
  V, Geerdens E, Clappier E, Porcu M, {\it et~al\/} (2013) Exome sequencing
  identifies mutation in {CNOT3 and ribosomal genes RPL5 and RPL10 in T-cell}
  acute lymphoblastic leukemia. {\it Nature genetics\/} {\bf 45}: 186--190

\bibitem[{Fabian and Sonenberg(2012)Fabian \&
  Sonenberg}]{sonenberg2012_microRNAs}
Fabian MR, Sonenberg N (2012) The mechanics of miRNA-mediated gene silencing: a
  look under the hood of miRISC. {\it Nature structural  molecular biology\/}
  {\bf 19}: 586--593

\bibitem[{Galkin {\it et~al\/}(2007)Galkin, Bentley, Gupta, Compton, Mazumder,
  Kinzy, Merrick, Hatzoglou, Pestova, Hellen {\it et~al\/}}]{galkin2007roles}
Galkin O, Bentley AA, Gupta S, Compton BA, Mazumder B, Kinzy TG, Merrick WC,
  Hatzoglou M, Pestova TV, Hellen CU, {\it et~al\/} (2007) {Roles of the
  negatively charged N-terminal extension of {\it Saccharomyces cerevisiae}
  ribosomal protein S5 revealed by characterization of a yeast strain
  containing human ribosomal protein S5}. {\it RNA\/} {\bf 13}: 2116--2128

\bibitem[{Gilbert(2011)}]{gilbert2011functional}
Gilbert WV (2011) Functional specialization of ribosomes? {\it Trends in
  biochemical sciences\/} {\bf 36}: 127--132

\bibitem[{Gingras {\it et~al\/}(1999)Gingras, Raught \&
  Sonenberg}]{Sonenberg_1999_eif4}
Gingras AC, Raught B, Sonenberg N (1999) {eIF4} initiation factors: effectors
  of mRNA recruitment to ribosomes and regulators of translation. {\it Annual
  review of biochemistry\/} {\bf 68}: 913--963

\bibitem[{Goodman and Rich(1963)Goodman \& Rich}]{goodman1963mechanism}
Goodman HM, Rich A (1963) Mechanism of polyribosome action during protein
  synthesis. {\it Nature\/} {\bf 199}: 318--322

\bibitem[{Hardy(1975)}]{hardy1975stoichiometry}
Hardy SJ (1975) The stoichiometry of the ribosomal proteins of {{\it
  Escherichia coli}}. {\it Molecular and General Genetics MGG\/} {\bf 140}:
  253--274

\bibitem[{Hendrickson {\it et~al\/}(2009)Hendrickson, Hogan, McCullough, Myers,
  Herschlag, Ferrell \& Brown}]{microRNAs_Ferrell_Brown_2009}
Hendrickson DG, Hogan DJ, McCullough HL, Myers JW, Herschlag D, Ferrell JE,
  Brown PO (2009) {Concordant regulation of translation and mRNA abundance for
  hundreds of targets of a human microRNA}. {\it PLoS biology\/} {\bf 7}:
  e1000238

\bibitem[{Hickman and Winston(2007)Hickman \& Winston}]{hickman2007heme}
Hickman M, Winston F (2007) Heme levels switch the function of {Hap1} of {{\it
  Saccharomyces cerevisiae}} between transcriptional activator and
  transcriptional repressor. {\it Molecular and Cellular Biology\/} {\bf 27}:
  7414--7424

\bibitem[{Horos {\it et~al\/}(2012)Horos, IJspeert, Pospisilova, Sendtner,
  Andrieu-Soler, Taskesen, Nieradka, Cmejla, Sendtner, Touw {\it
  et~al\/}}]{horos2012ribosomal}
Horos R, IJspeert H, Pospisilova D, Sendtner R, Andrieu-Soler C, Taskesen E,
  Nieradka A, Cmejla R, Sendtner M, Touw IP, {\it et~al\/} (2012) Ribosomal
  deficiencies in {Diamond--Blackfan} anemia impair translation of transcripts
  essential for differentiation of murine and human erythroblasts. {\it
  Blood\/} {\bf 119}: 262--272

\bibitem[{Jovanovic {\it et~al\/}(2015)Jovanovic, Rooney, Mertins, Przybylski,
  Chevrier, Satija, Rodriguez, Fields, Schwartz, Raychowdhury {\it
  et~al\/}}]{jovanovic2015dynamic}
Jovanovic M, Rooney MS, Mertins P, Przybylski D, Chevrier N, Satija R,
  Rodriguez EH, Fields AP, Schwartz S, Raychowdhury R, {\it et~al\/} (2015)
  Dynamic profiling of the protein life cycle in response to pathogens. {\it
  Science\/} {\bf 347}: 1259038

\bibitem[{Komili {\it et~al\/}(2007)Komili, Farny, Roth \&
  Silver}]{komili2007functional}
Komili S, Farny NG, Roth FP, Silver PA (2007) Functional specificity among
  ribosomal proteins regulates gene expression. {\it Cell\/} {\bf 131}:
  557--571

\bibitem[{Kondrashov {\it et~al\/}(2011)Kondrashov, Pusic, Stumpf, Shimizu,
  Hsieh, Xue, Ishijima, Shiroishi \& Barna}]{RPL38_kondrashov2011ribosome}
Kondrashov N, Pusic A, Stumpf CR, Shimizu K, Hsieh AC, Xue S, Ishijima J,
  Shiroishi T, Barna M (2011) Ribosome-mediated specificity in {Hox} {mRNA}
  translation and vertebrate tissue patterning. {\it Cell\/} {\bf 145}:
  383--397

\bibitem[{Kwon {\it et~al\/}(2013)Kwon, Yi, Eichelbaum, F{\"o}hr, Fischer, You,
  Castello, Krijgsveld, Hentze \& Kim}]{kwon2013rna}
Kwon SC, Yi H, Eichelbaum K, F{\"o}hr S, Fischer B, You KT, Castello A,
  Krijgsveld J, Hentze MW, Kim VN (2013) The {RNA-binding} protein repertoire
  of embryonic stem cells. {\it Nature structural  molecular biology\/}

\bibitem[{Lawrence {\it et~al\/}(2014)Lawrence, Stojanov, Mermel, Robinson,
  Garraway, Golub, Meyerson, Gabriel, Lander \& Getz}]{lawrence2014discovery}
Lawrence MS, Stojanov P, Mermel CH, Robinson JT, Garraway LA, Golub TR,
  Meyerson M, Gabriel SB, Lander ES, Getz G (2014) Discovery and saturation
  analysis of cancer genes across 21 tumour types. {\it Nature\/} {\bf 505}:
  495--501

\bibitem[{Lee {\it et~al\/}(2013)Lee, Burdeinick-Kerr \&
  Whelan}]{lee2013ribosome}
Lee ASY, Burdeinick-Kerr R, Whelan SP (2013) A ribosome-specialized translation
  initiation pathway is required for cap-dependent translation of vesicular
  stomatitis virus {mRNAs}. {\it Proceedings of the National Academy of
  Sciences\/} {\bf 110}: 324--329

\bibitem[{Mauro and Edelman(2002)Mauro \& Edelman}]{mauro2002ribosome_filter}
Mauro VP, Edelman GM (2002) The ribosome filter hypothesis. {\it Proceedings of
  the National Academy of Sciences\/} {\bf 99}: 12031--12036

\bibitem[{Mazumder {\it et~al\/}(2003)Mazumder, Sampath, Seshadri, Maitra,
  DiCorleto \& Fox}]{RPL13_Paul_Fox_2003}
Mazumder B, Sampath P, Seshadri V, Maitra RK, DiCorleto PE, Fox PL (2003)
  {Regulated release of L13a from the 60S ribosomal subunit as a mechanism of
  transcript-specific translational control}. {\it Cell\/} {\bf 115}: 187--198

\bibitem[{Noll {\it et~al\/}(1963)Noll, Staehelin \&
  Wettstein}]{noll1963ribosomal}
Noll H, Staehelin T, Wettstein F (1963) Ribosomal aggregates engaged in protein
  synthesis: ergosome breakdown and messenger ribonucleic acid transport. {\it
  Nature\/} {\bf 198}: 632--638

\bibitem[{O'Leary {\it et~al\/}(2013)O'Leary, Schreiber, Zhang, Duc, Rao, Hale,
  Academia, Shah, Morton, Holstein {\it et~al\/}}]{rpl22_repressing2013}
O'Leary MN, Schreiber KH, Zhang Y, Duc ACE, Rao S, Hale JS, Academia EC, Shah
  SR, Morton JF, Holstein CA, {\it et~al\/} (2013) The ribosomal protein
  {Rpl22} controls ribosome composition by directly repressing expression of
  its own paralog, {Rpl22l1}. {\it PLoS genetics\/} {\bf 9}: e1003708

\bibitem[{Ori {\it et~al\/}(2013)Ori, Banterle, Iskar, Andr{\'e}s-Pons, Escher,
  Khanh~Bui, Sparks, Solis-Mezarino, Rinner, Bork {\it
  et~al\/}}]{npc_2013_ori_beck}
Ori A, Banterle N, Iskar M, Andr{\'e}s-Pons A, Escher C, Khanh~Bui H, Sparks L,
  Solis-Mezarino V, Rinner O, Bork P, {\it et~al\/} (2013) Cell type-specific
  nuclear pores: a case in point for context-dependent stoichiometry of
  molecular machines. {\it Molecular systems biology\/} {\bf 9}

\bibitem[{Parenteau {\it et~al\/}(2011)Parenteau, Durand, Morin, Gagnon,
  Lucier, Wellinger, Chabot \& Abou~Elela}]{parenteau2011introns}
Parenteau J, Durand M, Morin G, Gagnon J, Lucier JF, Wellinger RJ, Chabot B,
  Abou~Elela S (2011) Introns within ribosomal protein genes regulate the
  production and function of yeast ribosomes. {\it Cell\/} {\bf 147}: 320--331

\bibitem[{Qian {\it et~al\/}(2012)Qian, Ma, Xiao, Wang \&
  Zhang}]{qian2012genomic}
Qian W, Ma D, Xiao C, Wang Z, Zhang J (2012) The genomic landscape and
  evolutionary resolution of antagonistic pleiotropy in yeast. {\it Cell
  Reports\/} {\bf 2}: 1399--1410

\bibitem[{Ramagopal(1990)}]{ramagopal1990induction}
Ramagopal S (1990) Induction of cell-specific ribosomal proteins in aggregation
  competent nonmorphogenetic {{\it Dictyostelium discoideum}}. {\it
  Biochemistry and Cell Biology\/} {\bf 68}: 1281--1287

\bibitem[{Ramagopal and Ennis(1981)Ramagopal \&
  Ennis}]{ramagopal1981regulation}
Ramagopal S, Ennis HL (1981) Regulation of synthesis of cell-specific ribosomal
  proteins during differentiation of {{\it Dictyostelium discoideum}}. {\it
  Proceedings of the National Academy of Sciences\/} {\bf 78}: 3083--3087

\bibitem[{Reschke {\it et~al\/}(2013)Reschke, Clohessy, Seitzer, Goldstein,
  Breitkopf, Schmolze, Ala, Asara, Beck \&
  Pandolfi}]{riboproteome_2013_Pandolfi}
Reschke M, Clohessy JG, Seitzer N, Goldstein DP, Breitkopf SB, Schmolze DB, Ala
  U, Asara JM, Beck AH, Pandolfi PP (2013) Characterization and Analysis of the
  Composition and Dynamics of the Mammalian Riboproteome. {\it Cell Reports\/}
  {\bf 4}: 1276--1287

\bibitem[{Shalem {\it et~al\/}(2014)Shalem, Sanjana, Hartenian, Shi, Scott,
  Mikkelsen, Heckl, Ebert, Root, Doench {\it et~al\/}}]{shalem2014genome}
Shalem O, Sanjana NE, Hartenian E, Shi X, Scott DA, Mikkelsen TS, Heckl D,
  Ebert BL, Root DE, Doench JG, {\it et~al\/} (2014) Genome-scale {CRISPR-Cas9}
  knockout screening in human cells. {\it Science\/} {\bf 343}: 84--87

\bibitem[{Slavov {\it et~al\/}(2012)Slavov, Airoldi, van Oudenaarden \&
  Botstein}]{Slavov_emc}
Slavov N, Airoldi EM, van Oudenaarden A, Botstein D (2012) A conserved cell
  growth cycle can account for the environmental stress responses of divergent
  eukaryotes. {\it Molecular Biology of the Cell\/} {\bf 23}: 1986 -- 1997

\bibitem[{Slavov and Botstein(2011)Slavov \& Botstein}]{Slavov_eth_grr}
Slavov N, Botstein D (2011) Coupling among growth rate response, metabolic
  cycle, and cell division cycle in yeast. {\it Molecular Biology of the
  Cell\/} {\bf 22}: 1997--2009

\bibitem[{Slavov {\it et~al\/}(2014)Slavov, Budnik, Schwab, Airoldi \& van
  Oudenaarden}]{Slavov_exp}
Slavov N, Budnik B, Schwab D, Airoldi E, van Oudenaarden A (2014) {Constant
  Growth Rate Can Be Supported by Decreasing Energy Flux and Increasing Aerobic
  Glycolysis}. {\it Cell Reports\/} {\bf 7}: 705 -- 714

\bibitem[{Slavov and Dawson(2009)Slavov \& Dawson}]{Slavov_2009}
Slavov N, Dawson KA (2009) {Correlation signature of the macroscopic states of
  the gene regulatory network in cancer}. {\it Proceedings of the National
  Academy of Sciences\/} {\bf 106}: 4079--4084

\bibitem[{Sonenberg and Hinnebusch(2009)Sonenberg \&
  Hinnebusch}]{sonenberg2009regulation}
Sonenberg N, Hinnebusch AG (2009) Regulation of translation initiation in
  eukaryotes: mechanisms and biological targets. {\it Cell\/} {\bf 136}:
  731--745

\bibitem[{Tiruneh {\it et~al\/}(2013)Tiruneh, Kim, Gallie, Roy \& von
  Arnim}]{tiruneh2013global}
Tiruneh BS, Kim BH, Gallie DR, Roy B, von Arnim AG (2013) The global
  translation profile in a ribosomal protein mutant resembles that of an {eIF3}
  mutant. {\it BMC biology\/} {\bf 11}: 123

\bibitem[{Topisirovic and Sonenberg(2011)Topisirovic \&
  Sonenberg}]{Sonenberg2011Preview}
Topisirovic I, Sonenberg N (2011) Translational control by the eukaryotic
  ribosome. {\it Cell\/} {\bf 145}: 333--334

\bibitem[{Vaidyanathan {\it et~al\/}(2014)Vaidyanathan, Zinshteyn, Thompson \&
  Gilbert}]{pavan2014wendy}
Vaidyanathan PP, Zinshteyn B, Thompson MK, Gilbert WV (2014) Protein kinase {A}
  regulates gene-specific translational adaptation in differentiating yeast.
  {\it RNA\/} : 10.1261/rna.044552.114

\bibitem[{Wang {\it et~al\/}(2007)Wang, Chen, Baker, Chen, Kaiser \&
  Huang}]{Huang_2007_proteasome}
Wang X, Chen CF, Baker PR, Chen Pl, Kaiser P, Huang L (2007) {Mass
  spectrometric characterization of the affinity-purified human 26S proteasome
  complex}. {\it Biochemistry\/} {\bf 46}: 3553--3565

\bibitem[{Warner(1999)}]{warner1999economics}
Warner JR (1999) The economics of ribosome biosynthesis in yeast. {\it Trends
  in biochemical sciences\/} {\bf 24}: 437--440

\bibitem[{Warner {\it et~al\/}(1963)Warner, Knopf \& Rich}]{warner1963multiple}
Warner JR, Knopf PM, Rich A (1963) A multiple ribosomal structure in protein
  synthesis. {\it Proceedings of the National Academy of Sciences of the United
  States of America\/} {\bf 49}: 122

\bibitem[{Warner and McIntosh(2009)Warner \& McIntosh}]{warner2009common}
Warner JR, McIntosh KB (2009) How common are extraribosomal functions of
  ribosomal proteins? {\it Molecular cell\/} {\bf 34}: 3--11

\bibitem[{Weber(1972)}]{weber1972stoichiometric}
Weber HJ (1972) Stoichiometric measurements of {30S} and {50S} ribosomal
  proteins from {{\it Escherichia coli}}. {\it Molecular and General Genetics
  MGG\/} {\bf 119}: 233--248

\bibitem[{Westermann {\it et~al\/}(1976)Westermann, Heumann \&
  Bielka}]{westermann1976stoichiometry}
Westermann P, Heumann W, Bielka H (1976) On the stoichiometry of proteins in
  the small ribosomal subunit of hepatoma ascites cells. {\it FEBS letters\/}
  {\bf 62}: 132--135

\bibitem[{Wettstein {\it et~al\/}(1963)Wettstein, Staehelin \&
  Noll}]{wettstein1963ribosomal}
Wettstein F, Staehelin T, Noll H (1963) Ribosomal aggregate engaged in protein
  synthesis: characterization of the ergosome. {\it Nature\/} {\bf 197}:
  430--435

\bibitem[{Wool(1996)}]{wool1996extraribosomal}
Wool IG (1996) Extraribosomal functions of ribosomal proteins. {\it Trends in
  biochemical sciences\/} {\bf 21}: 164--165

\bibitem[{Xue and Barna(2012)Xue \& Barna}]{xue2012specialized}
Xue S, Barna M (2012) Specialized ribosomes: a new frontier in gene regulation
  and organismal biology. {\it Nature Reviews Molecular Cell Biology\/} {\bf
  13}: 355--369

\bibitem[{Xue {\it et~al\/}(2015)Xue, Tian, Fujii, Kladwang, Das \&
  Barna}]{xue2015rna}
Xue S, Tian S, Fujii K, Kladwang W, Das R, Barna M (2015) {RNA regulons in Hox
  5'-UTRs confer ribosome specificity to gene regulation}. {\it Nature\/} {\bf
  517}: 33--38

\bibitem[{Young and Bremer(1976)Young \& Bremer}]{young1976polypeptide}
Young R, Bremer H (1976) Polypeptide-chain-elongation rate in {{\it Escherichia
  coli}} {B/r} as a function of growth rate. {\it Biochem J\/} {\bf 160}:
  185--194

\end{thebibliography}
\end{document}